# Reduced-Order Multiscale Modeling of Plastic Deformations in 3D Alloys with Spatially Varying Porosity by Deflated Clustering Analysis

Shiguang Deng [a], Carl Soderhjelm [a], Diran Apelian [a], Ramin Bostanabad [b, *]

[a] ACRC, Materials Science and Engineering, University of California, Irvine, CA, USA
[b] Mechanical and Aerospace Engineering, University of California, Irvine, CA, USA

**Abstract**

Aluminum alloys are increasingly utilized as lightweight materials in the automobile industry due to their superior capability in withstanding high mechanical loads. A significant challenge impeding the large-scale use of these alloys in high-performance applications is the presence of manufacturing-induced, spatially varying porosity defects. In order to understand the impacts of these defects on the macro-mechanical properties of cast alloys, multiscale simulations are often required. In this paper, we introduce a computationally efficient reduced-order multiscale framework to simulate the behavior of metallic components containing process-induced porosity under irreversible nonlinear deformations. In our approach, we start with a data compression scheme that significantly reduces the number of unknown macroscale and microscale variables by agglomerating close-by finite element nodes into a limited number of clusters. Then, we use deflation methods to project these variables into a lower-dimensional space where the material's elastoplastic behaviors are approximated. Finally, we solve for the unknown variables and map them back to the original, high-dimensional space. We call our method deflated clustering analysis and by comparing it to direct numerical simulations we demonstrate that it accurately captures macroscale deformations and microscopic effective responses. To illustrate the effect of microscale pores on the macroscopic response of a cast component, we conduct multi-scale simulations with spatially varying local heterogeneities that are modeled with a microstructure characterization and reconstruction algorithm.

## 1. Introduction

Cast aluminum alloys are heavily used in industrial applications where they are typically subject to plastic deformation to fully exploit their load-carrying capacity. These alloys have a heterogeneous nature which is primarily due to process-induced defects and variations. Pores are one of the most critical defects in cast metals; they possess spatially varying morphology and distribution (see Figure 1) and are generally due to gas or shrinkage [1,2]. Since pores considerably impact the performance of cast alloys [3,4], it is crucial to quantify their effects on the mechanical performance of a macrostructure subject to path-dependent plastic deformations. This quantification is typically achieved via multiscale simulations because pores are much smaller than cast components. While classic multiscale simulations resolve all the fine microstructural details, they are memory intensive and computationally demanding. To address this issue, we propose a computationally efficient reduce-order multiscale model to simulate the elastoplastic behaviors of cast alloys with process-induced, spatially varying porosity defects. In our approach, we employ a clustering-based domain decomposition that universally applies to macro- and micro- domains to accelerate, respectively, high-fidelity calculation of macroscale deformations and effective microscopic responses.

---

* Corresponding author.
   E-mail address: raminb@uci.edu (R. Bostanabad).

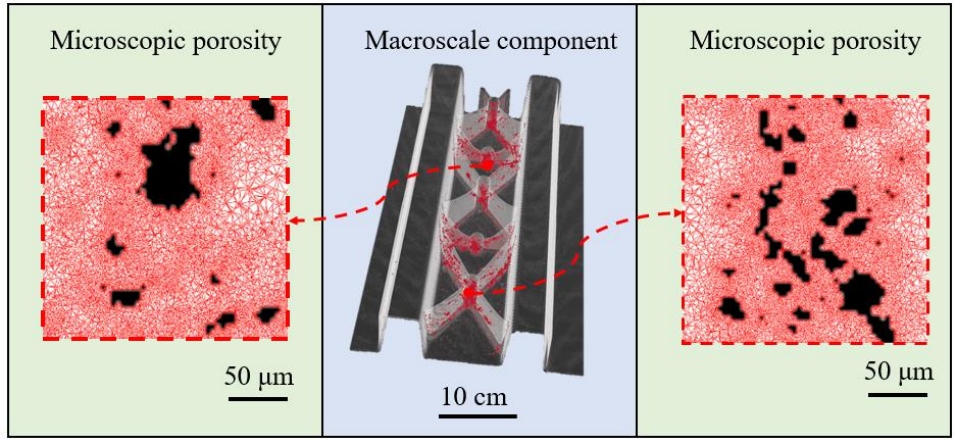

**Figure 1 Spatial microstructure variations:** An aluminum W-profile plate manufactured via high pressure die casting [5]. The plate has more than 1300 micro-pores whose spatial distribution and morphology are reconstructed from 3D X-ray tomography. This plate is used as a supportive structure in automobiles and the existence of pores significantly impairs its mechanical performance when subject to elastoplastic deformations. To model the pore morphology, extremely fine mesh is needed in the vicinity of pore boundaries.

Traditional phenomenological material models [5,6] formulate the mean behavior of materials and fail to capture highly localized microstructure-dependent deformations. They are also problem- and material-dependent and require calibration against experiments. These drawbacks can be overcome via computational homogenization which is a well-established and popular method for multiscale modeling that involves the solution of two (nested) boundary value problems (BVPs) that characterize the macroscopic and microscopic deformations. Assuming the characteristic microstructural length-scale is much smaller than the macrostructural size and load variability, each iteration of a first-order homogenization scheme starts by calculating the macroscopic deformation (gradient) tensor for every material point (aka integration or Gauss point) in the macrostructure[†]. This tensor is then used to construct the BVP that formulates the deformation of the unique microstructure assigned to the corresponding macroscale material point. Once the microstructural BVPs are solved, the macroscopic stress tensors are obtained by volume-averaging the corresponding microstructural stress fields. This iterative process is continued until equilibrium is achieved at both scales. This approach is also called $FE^2$ if the finite element method (FEM) is used to solve the BVPs at both scales, see Figure 2(a).

First-order homogenization scheme is a versatile strategy to model the macroscopic mechanical response of non-linear, multi-phase, multi-scale materials because it does not place any constitutive assumption on the overall material behavior at the microscale. However, it is computationally expensive and hence not suitable for large macroscale simulations whose microstructures spatially vary and include intricate details. In order to decrease computational costs while maintaining high accuracy and versatility, reduced-order models (ROMs) are developed. ROMs typically strike a balance between accuracy and cost by reducing the number of unknown variables and conducting offline simulations that accelerate online calculations. Some notable ROMs are based on the fast Fourier transformation (FFT) [8], spectral methods [9], principle component analysis (PCA) [10], proper generalized decomposition (PGD) [11], transformation field analysis (TFA) [12], nonuniform transformation field analysis (NTFA) [16,17], proper orthogonal decomposition (POD) [15], and self-consistent clustering analysis (SCA) [19–21].

---

[†] Since only the first gradient of the macroscopic displacement field is used, the method is called first order.

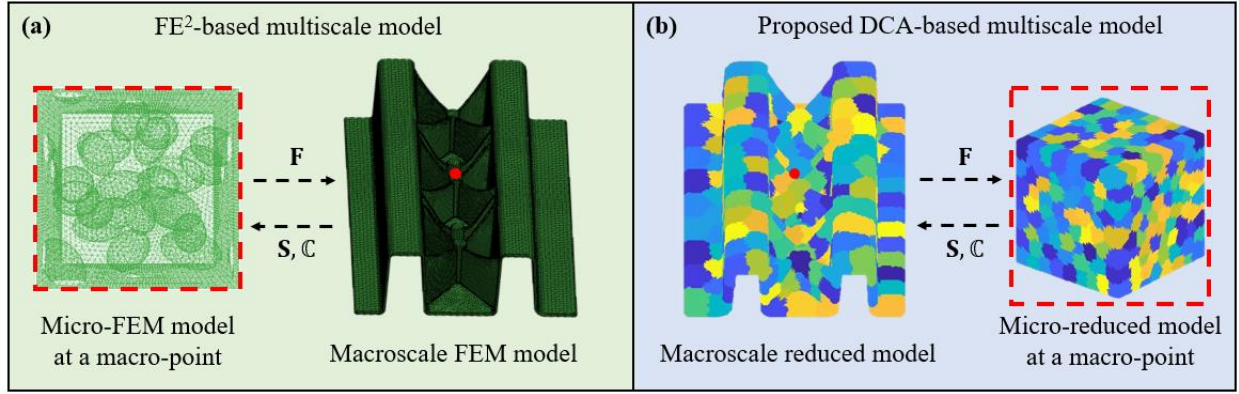

**Figure 2 Illustration of classic and proposed reduced-order multiscale models: (a)** First order computational homogenization via FE[2] where each macroscale material point is associated with a porous microstructure which spatially varies in realistic components. **(b)** The proposed deflated clustering analyses (DCA)-based multiscale model where computational efficiencies at both scales are significantly improved via a data compression algorithm that agglomerates close-by nodes into clusters, marked by different colors. In concurrent multiscale simulations, deformation gradients (**F**) are passed down from macroscale to microscale, and the homogenized stress (**S**) and tangent moduli ($\mathbb{C}$) are passed back.

The TFA method approximates the stress and strain field as uniform in each phase of a heterogeneous microstructure. The constituents are assumed as generalized standard materials [19] and the state of material points is defined by a vector of internal state variables associated with dissipative phenomena such as plasticity and damage. The evolution of state variables is controlled by analytical functions which involve thermodynamic forces and potentials [20]. TFA reduces the number of state variables to achieve high efficiency by expressing strain fields as a linear combination of truncated uniform eigenstrains. The NTFA method extends TFA by allowing each phase to possess spatially nonuniform fields constructed from incompressible and orthonormal eigenstrains. While the eigenstrains can be determined by numerical simulations in an offline stage where characteristic loads are applied on microstructures, a more efficient approach is to select eigenstrains through the POD procedure which considers the predetermined eigenstrains as a collection of samples. A small number of suitable eigenstrains can be extracted by minimizing the difference between the pre-recorded displacements and the ones constructed from the samples.

The SCA method is a recent ROM that consists of two primary stages. In an offline stage, microstructures are loaded with characteristic forces to determine the elastic responses at each material point. Points with similar strain concentration tensors are grouped into clusters where the stress and strain fields are assumed uniform. Cluster-to-cluster interactions are accounted for by the Green function in the discretized Lippmann-Schwinger equation, which represents the influence of the stress in one cluster onto the strain in another cluster. Since the online performance of the discretized Lippmann-Schwinger equation depends on the choice of reference materials, the isotropic linear elastic reference materials are corrected by a self-consistent scheme in the online stage that incrementally updates the reference material property to approximate the macroscopic modulus.

To reduce the computational costs related to the nested multiscale computations, an alternative approach is to develop decoupled methods [21]. In such methods, extensive microscale simulations are computed beforehand via either (single-scale) direct numerical simulation (DNS) or ROM, and the computed effective stress and strain fields are related by a surrogate model. The trained surrogate model works as a data-driven microstructural constitutive law in the online stage and provides homogenized responses to macroscale inquiries [22]. The online computation does not need to trace yield surfaces, nor does it require the nontrivial definition of flow rules. Instead, it necessitates extensive numerical simulations on various microstructures with different deformation paths to enable the surrogate model to learn the irreversible and path-dependent plastic phenomena directly from simulation data. In this approach, as most computational costs are allocated at generating sample data in the offline stage, the online computation is highly efficient since it only involves simple inquiry on predefined mapping functions such as kernel methods [23] and artificial neural networks (ANNs) [27,28].

To quantitatively investigate the influence of porosity defects on metal structure behaviors, pores have been incorporated into analysis models in many studies. Most analyses are based on DNS or $FE^2$ where pores are modeled with either simplified shapes or actual morphologies obtained from (non-destructive) inspections. For example, characteristic pore geometries are reconstructed from high-resolution tomography characterization in [26]. Their influence on elastoplastic behaviors is estimated by DNS that directly incorporates the reconstructed pore model in a microscale unit cell. In [27], the morphology of cast pores is identified by light microscopy in a fatigue crack initiation study where linear elastic studies via DNS are utilized to correlate pore shapes with local stress concentrations. In [28], heterogeneously distributed cast pores are reconstructed from a stack of microstructural serial images in a DNS-based micromechanics model where local fields are significantly influenced by pore geometrical features such as size, orientation, and spatial arrangement. A significant challenge of integrating actual pore morphology into analysis models is discretization or meshing: small and irregularly shaped elements are generated in the vicinity of pores which substantially increases element numbers, deteriorates mesh quality, and slows solver convergence rate. An alternative strategy is to simplify pore geometries to improve mesh quality and reduce pore shape descriptors. For instance, spherical voids are used in [29] to investigate the critical sizes of pores for crack nucleation under dynamic loading conditions in an additive manufactured nickel-based alloy where pores are explicitly added to microstructural DNS models with varying size, location, and spatial proximity. Ellipsoid voids are used in [30] where microstructural damage is propagated via an adaptive multiscale model whose inter-scale coupling is based on asymptotic homogenization in an $FE^2$ framework. In [31], an $FE^2$-based sequential multiscale model is developed to consider the influence of cast porosity on the plastic behaviors of a nickel-based superalloy. In this study, pores are approximated by intersecting three identical ellipsoids at geometric centers to account for convex-concave geometries, and the pore model is calibrated so that its volume and sphericity are consistent with the actual pores observed by X-ray.

Despite these advances to incorporate the influence of micro-porosity defects on macro-structural performance, the following research gaps exist:
- Previous studies are primarily based on single-scale DNS or $FE^2$. DNS with reconstructed pore morphologies often has a slow convergence rate (due to fine meshes with ill-shaped elements around the pores), and approaches based on $FE^2$ are generally memory intensive and computationally expensive. There is a need for a computationally efficient porosity analysis model that efficiently quantifies the effect of local heterogeneities on macroscale behavior.

- Most existing ROMs and surrogates require extensive exploration of the deformation space to collect sufficient samples in the offline stage when materials are subject to irreversible plastic deformations (as in TFA, POD, and ANN). Finding proper macro-constitutive equations and calibrating them against experiments (as in NTFA) is quite difficult. A new ROM which avoids these shortcomings is needed.
- Most synthetic porosity models oversimplify pore morphology. They are either two-dimensional (2D) or incorporate pore characteristics (especially its spatial distribution) in a heuristic manner. A more realistic three-dimensional (3D) porosity representation and analysis model is necessary.

We propose a novel mechanistic multiscale ROM coined as deflated clustering analysis (DCA) for simulating 3D heterogeneous alloys subject to elastoplastic deformations. The numerical advantages of DCA over classic multiscale models are demonstrated in Figure 2(b), where simulations at both macro- and micro- scales are accelerated by reduced models by systematically agglomerating close-by nodes into clusters. The proposed method projects solution variables into a lower-dimensional space for nonlinear simulations. It avoids extensive offline exploration and does not need empirical constitutive equations. Our ROM reduces computational costs by more than one order of magnitude without significant accuracy loss. Additionally, we develop a porosity-oriented microstructure characterization and reconstruction algorithm to associate spatially varying microstructures with a macro-component and, in turn, study the effects of porosity and its spatial distribution on the component performance.

The remainder of the paper is organized as follows. Section 2 reviews the first-order computational homogenization theory which serves as the foundation of our accelerated multiscale modeling approach. Section 3 describes the proposed DCA framework which is augmented with a porosity-oriented microstructure characterization and reconstruction algorithm detailed in Section 4. In Section 5, the efficiency and accuracy of our method are evaluated via a wide range of numerical experiments. Conclusion and future works are provided in Section 6.

## 2. First-order computational homogenization

Computational homogenization aims to approximate the effective response of a representative volume element (RVE) of a microstructure and assumes that different spatial scales in a generic material can be identified and distinguished. More specifically, this method presumes that each macroscopic material point is associated with an RVE which satisfies the scale separation principle. That is, the average size ($l_\mu$) of material heterogeneities is much smaller than the characterize size of the RVE ($l_\mathrm{m}$) which is itself significantly smaller than the characteristic length of the macrostructure ($l_\mathrm{M}$):

$$l_\mu \ll l_\mathrm{m} \ll l_\mathrm{M} \tag{1}$$

where the subscripts 'M' and 'm' denote macroscale and microscale. In this work, the scale separation assumption is satisfied since the microscale pores are considerably smaller than the RVEs which are much smaller than cast components.

In what follows, vectors are written in bold lower case while tensors are typed in bold uppercase.

## 2.1. Macroscale problem

In an infinitesimal deformation framework, if a macrostructure is in a quasi-static state, its equilibrium equation in the weak form [32] reads as:

$$\int_{\Omega_{0M}} \left[ \mathbf{S_M}(\mathbf{X},t) : \nabla_0 \boldsymbol{\eta} - \mathbf{b_M} \boldsymbol{\eta} \right] dV - \int_{\Gamma_{0M}} \mathbf{\bar{t}_M} \boldsymbol{\eta} dA = 0 \quad \forall \boldsymbol{\eta} \in \Psi \quad (2)$$

where $\mathbf{S_M}(\mathbf{X}, t)$ is the unknown macroscopic stress tensor at a generic macroscopic material point $\mathbf{X}$ and at any instance in time $t$, $\mathbf{b_M}$ and $\mathbf{\bar{t}_M}$ are, respectively, the given body force per unit volume on the undeformed domain $\Omega_{0M}$ and the external traction force per unit area on the undeformed domain surface $\Gamma_{0M}$, $\boldsymbol{\eta}$ represents an admissible virtual displacement field in the space of virtual displacement $\Psi$, and $\nabla_0$ is the gradient operator with respect to the reference configuration. The symbol ':' represents the double dot product in the tensor notation, which denotes the contraction of a pair of repeated indices that appear in the same order of the two multiplying tensors.

The equilibrium equation can be written in the strong form as a BVP [32] as:

$$\mathbf{S_M}(\mathbf{X},t) \cdot \nabla_0 + \mathbf{b_M} = \mathbf{0} \quad \forall \mathbf{X} \in \Omega_{0M} \quad (3)$$

$$\mathbf{u_M}(\mathbf{X},t) = \mathbf{\bar{u}_M} \quad \forall \mathbf{X} \in \Gamma_{0M}^{D} \quad (4)$$

$$\mathbf{S_M}(\mathbf{X},t) \cdot \mathbf{n_M} = \mathbf{\bar{t}_M} \quad \forall \mathbf{X} \in \Gamma_{0M}^{N} \quad (5)$$

where $\mathbf{u_M}$ indicates the unknown displacement variables, $\mathbf{\bar{u}_M}$ is the prescribed displacement field on the Dirichlet boundary $\Gamma_{0M}^{D}$, $\mathbf{\bar{t}_M}$ represents the surface traction over the Neumann boundary $\Gamma_{0M}^{N}$, and $\mathbf{n_M}$ denotes the outward unit vector to the boundary of the undeformed macrostructural domain $\Omega_{0M}$.

## 2.2. Microscale problem

The displacement field at the microscale is decomposed into two parts [33]:

$$\mathbf{u_m}(\mathbf{x},t) = \left[ \mathbf{F_M}(\mathbf{X},t) - \mathbf{I} \right](\mathbf{x} - \mathbf{x_0}) + \mathbf{\tilde{u}}(\mathbf{x},t) \quad \forall \mathbf{x} \in \Omega_{0m} \quad (6)$$

where $\mathbf{u_m}(\mathbf{x}, t)$ indicates the unknown displacement at an arbitrary point $\mathbf{x}$ in a microstructure at time $t$, $\mathbf{F_M}(\mathbf{X}, t)$ is the macroscopic deformation gradient at the macroscopic point $\mathbf{X}$ which corresponds to the microstructure $\Omega_{0m}$, $\mathbf{I}$ is the identity matrix, and $\mathbf{x_0}$ represents an arbitrary reference point in the microstructure. The first term on the right-hand side of Equation (6) represents the homogeneous deformation given by the macroscopic deformation gradient, and the second term indicates a microscopic displacement fluctuation field $\mathbf{\tilde{u}}$.

The weak form of the microscale equilibrium equation in the absence of dynamics is [34]:

$$\int_{\Omega_{0m}} \mathbf{S_m}(\mathbf{x},t) : \nabla_0 \boldsymbol{\eta} dV - \int_{\Gamma_{0m}} \mathbf{\bar{t}_m} \boldsymbol{\eta} dA = 0 \quad \forall \boldsymbol{\eta} \in \Psi \quad (7)$$

where $\mathbf{S_m}(\mathbf{x}, t)$ is the microscale stress tensor, $\mathbf{\bar{t}_m}$ is the external traction force on the reference microstructural domain $\Omega_{0m}$ and surface $\Gamma_{0m}$. Similar to the macroscale problem, the microscale equilibrium equation can be written in the strong form:

$$\mathbf{S_m}(\mathbf{x},t) \cdot \nabla_0 = \mathbf{0} \quad \forall \mathbf{x} \in \Omega_{0m} \quad (8)$$

$$\mathbf{S_m}(\mathbf{x},t) \cdot \mathbf{n_m} = \mathbf{\bar{t}_m} \quad \forall \mathbf{x} \in \Gamma_{0m}^{N} \quad (9)$$

where $\bar{\mathbf{t}}_\mathbf{m}$ is the given surface traction per unit area on the boundary $\Gamma_{0m}^N$ of reference microscale domain with the outward unit normal vector $\mathbf{n_m}$.

*2.3. Scale transition*

In the context of computational homogenization, scale coupling is established by volume averaging [21]. Specifically, the macro-to-micro transition is formulated via kinematic averaging where the deformation gradient of a generic macroscopic point at a given time equals to the volume average of its microscopic counterpart:

$$\mathbf{F_M}(\mathbf{X},t) = \frac{1}{V_{0m}} \int_{\Omega_{0m}} \mathbf{F_m}(\mathbf{x},t) dV \qquad \forall \mathbf{x} \in \Omega_{0m} \tag{10}$$

in which $V_{0m}$ represents the volume of the undeformed micro-domain $\Omega_{0m}$.

With the definition of microscale displacement in Equation (6) and the kinematic scale transition in Equation (10), one can define the minimal kinematic admissibility constraint [35] as:

$$\int_{\Gamma_{0m}} \tilde{\mathbf{u}}(\mathbf{x},t) \otimes \mathbf{n_m} dA = 0 \tag{11}$$

The boundary conditions on the microstructure should be chosen such that the left-hand side (LHS) of Equation (11) vanishes due to the contribution from the microscopic displacement fluctuation field. Boundary conditions in this category are called admissible kinematic boundary conditions and some of the most commonly used ones include: minimal kinematic boundary conditions, which only need the LHS of Equation (11) to vanish in an integrated manner, Taylor assumption which does not allow any displacement fluctuations within micro-domain as in Equation (12), uniform displacement boundary condition which explicitly prescribes displacements on domain boundaries as in Equation (13), and periodic boundary conditions which require periodic micro-fluctuations on the corresponding points at the opposite boundary surfaces as in Equations (14) and (15).

$$\tilde{\mathbf{u}}_\mathbf{m}(\mathbf{x},t) = \mathbf{0} \qquad \forall \mathbf{x} \in \Omega_{0m} \tag{12}$$

$$\tilde{\mathbf{u}}_\mathbf{m}(\mathbf{x},t) = \mathbf{0} \qquad \forall \mathbf{x} \in \Gamma_{0m} \tag{13}$$

$$\tilde{\mathbf{u}}_\mathbf{m}(\mathbf{x}^+,t) = \tilde{\mathbf{u}}_\mathbf{m}(\mathbf{x}^-,t) \qquad \forall \mathbf{x}^+ \in \Gamma_{0m}^+, \; \forall \mathbf{x}^- \in \Gamma_{0m}^- \tag{14}$$

$$\bar{\mathbf{t}}_\mathbf{m}(\mathbf{x}^+,t) = -\bar{\mathbf{t}}_\mathbf{m}(\mathbf{x}^-,t) \qquad \forall \mathbf{x}^+ \in \Gamma_{0m}^+, \; \forall \mathbf{x}^- \in \Gamma_{0m}^- \tag{15}$$

In Equation (14) and (15), the micro-domain boundary $\Gamma_{0m}$ is divided into positive ($\Gamma_{0m}^+$) and negative ($\Gamma_{0m}^-$) parts where for each point $\mathbf{x}^+$ residing on the positive part there is a corresponding point $\mathbf{x}^-$ on the negative part. In this work, we adopt the uniform displacement boundary condition.

Scale transition from microscale to macroscale is based on the Hill-Mandel condition [36], which requires the macroscale stress power to equal the volume average of its microscopic counterpart over the micro-domain. Formulated in terms of a work conjugated set, the Hill-Mandel condition reads:

$$\frac{1}{V_{0m}} \int_{\Omega_{0m}} \mathbf{S_m}(\mathbf{x},t) : \delta \mathbf{E_m}(\mathbf{x},t) dV = \mathbf{S_M}(\mathbf{X},t) : \delta \mathbf{E_M}(\mathbf{X},t) \tag{16}$$

where $\mathbf{E_m}(\mathbf{x},t)$ and $\mathbf{E_M}(\mathbf{X},t)$ are microscale and macroscale strain tensors, respectively.

Based on the energy consistency the macroscale stress is expressed as the volume average of its microscale counterpart by applying the admissible kinematic boundary conditions on microstructural boundaries:

$$\mathbf{S_M}(\mathbf{X},t) = \frac{1}{V_{0m}} \int_{\Omega_{0m}} \mathbf{S_m}(\mathbf{x},t) dV \tag{17}$$

One can use Equation (17) to numerically compute the homogenized stress tensor over the undeformed micro-domain. Alternatively, the macroscopic stress can be more efficiently computed:

$$\mathbf{S_M}(\mathbf{X},t) = \frac{1}{V_{0m}} \int_{\Gamma_{0m}} \bar{\mathbf{t}}_\mathbf{m} \otimes (\mathbf{x} - \mathbf{x_0}) dA \tag{18}$$

By expressing the microscale displacement in terms of micro-fluctuations in Equation (6), the Hill-Mandel condition is simplified as:

$$\int_{\Omega_{0m}} \mathbf{b_m} \boldsymbol{\eta} dV = 0 \quad \forall \boldsymbol{\eta} \in \Psi \tag{19}$$

$$\int_{\Gamma_{0m}^N} \bar{\mathbf{t}}_\mathbf{m} \boldsymbol{\eta} dA = 0 \quad \forall \boldsymbol{\eta} \in \Psi \tag{20}$$

where it is demonstrated in [34] that the microscopic body force $\mathbf{b_m}$ and surface traction $\bar{\mathbf{t}}_\mathbf{m}$ are essentially the reaction forces for the displacement fluctuations due to the imposed kinematic constraint on microstructural boundaries.

## 2.4. Homogenized material moduli

Although no explicit constitutive information is available at the macroscale, the macroscopic tangent stiffness is often required in multiscale simulations. The tangent stiffness can be numerically evaluated using the relation between the variations of the stress and deformation at each macroscale point. Conventional strategies use direct numerical differentiation of macroscopic stress-strain relation [37]. A more efficient alternative is the condensation method [38] which starts by partitioning the microscale system of equations as:

$$\begin{bmatrix} \mathbf{K_{pp}} & \mathbf{K_{pf}} \\ \mathbf{K_{fp}} & \mathbf{K_{ff}} \end{bmatrix} \begin{bmatrix} \delta \mathbf{u_p} \\ \delta \mathbf{u_f} \end{bmatrix} = \begin{bmatrix} \delta \mathbf{f_p} \\ \mathbf{0} \end{bmatrix} \tag{21}$$

where $\delta \mathbf{u_p}$ and $\delta \mathbf{u_f}$ are the displacement variations at the prescribed and free nodes in the microstructure, and $\delta \mathbf{f_p}$ is the external force on the prescribed nodes. $\mathbf{K_{pp}}, \mathbf{K_{pf}}, \mathbf{K_{fp}}$ and $\mathbf{K_{ff}}$ represent the corresponding partitions of the microstructural stiffness matrix. Eliminating $\delta \mathbf{u_f}$ from Equation (21) leads to a reduced system that directly relates the variations of the prescribed displacements with nodal forces:

$$\mathbf{K_r} \delta \mathbf{u_p} = \delta \mathbf{f_p} \tag{22}$$

$$\mathbf{K_r} = \mathbf{K_{pp}} - \mathbf{K_{pf}} (\mathbf{K_{ff}})^{-1} \mathbf{K_{fp}} \tag{23}$$

By substituting the variation of the nodal force from Equation (22) into the variational macroscopic stress in Equation (18), the macroscopic fourth-order consistent tangent moduli $\mathbb{C}$ can be derived as:

$$\delta \mathbf{S_M}(\mathbf{X},t) = \mathbb{C} : \delta \mathbf{E_M^T}(\mathbf{X},t) \tag{24}$$

$$\mathbb{C} = \frac{1}{V_{0m}} \left[ (\mathbf{x} - \mathbf{x_0}) \otimes \mathbf{K_r} \otimes (\mathbf{x} - \mathbf{x_0}) \right]^{LT} \qquad (25)$$

where the superscript 'LT' denotes the transposition between the two left indices. Readers are referred to [38] for more details.

## 3. Proposed framework of deflated clustering analysis

Integration of local porosity characteristics with homogenization-based multiscale models (e.g., FE$^2$) is challenging primarily because: (1) capturing detailed pore morphologies requires a fine mesh and hence a large number of degrees of freedom (DOF), and (2) nonlinear microstructural computations that are embedded at every macro-material point are expensive. Our proposed DCA addresses these two challenges while maintaining sufficiently high accuracy.

The proposed ROM includes an acceleration scheme for each length-scale: (1) we adopt an incremental deflation method to accelerate macroscale simulations while the deformation gradients at each macro-integration point are computed with no accuracy loss, and (2) we propose a microscopic projection method to speed up the iterative elastoplastic solution process in the microstructures with high-fidelity homogenized responses. Both acceleration schemes rely on the spatial domain decomposition technique discussed in Section 3.1. The macroscale incremental deflation method and the microscopic projection ROM are discussed in Section 3.2 and Section 3.3, respectively.

*3.1. Spatial domain decomposition*

Spatial domain decomposition converts a specific micro- or macro- structural domain into a set of interactive clusters with irregular shapes and distinct sizes. It can be performed in different ways. For instance, in [19,21], material points are grouped based on their mechanical behavior. To determine the grouping metric, six orthogonal loadings are performed within the elastic regime to compute the 36 independent components of strain concentration tensor at each material point. Then, the points with similar strain concentration tensors are grouped into a distinct cluster. An alternative approach is to group points by their spatial proximity [39] where, e.g., an existing mesh is divided by multiple predefined bounding boxes and then all nodes in the same box are lumped into one cluster.

In this work, we implement domain decomposition based on nodal spatial positions where we first record every node's coordinate and then agglomerate nodes with similar nodal coordinates using a clustering algorithm. While there are different clustering methods such as k-means [40] and Voronoi diagram [41], k-means clustering is adopted here due to its robustness and ease of use. Specifically, we use the 'kmeans' function in MATLAB [42] to partition the mesh nodes into $k$ predefined clusters. A specific node is assigned to the cluster whose centroid has the closest distance to its nodal coordinates. During this assignment, the cluster shapes are iteratively updated by including the nodes that minimize the within-cluster variance in terms of squared Euclidean distance for the $k$ sets $\mathbf{S} = \{S^1, S^2, \dots, S^k\}$ as:

$$\mathbf{S} = \arg\min_{\mathbf{S'}} \sum_{I=1}^{k} \sum_{n \in S^I} \left\| \mathbf{X_n} - \bar{\mathbf{X}}_\mathbf{I} \right\|^2 \qquad (26)$$

in which $\mathbf{X_n}$ is the nodal coordinates of the $n^{th}$ node, and $\bar{X}_I$ is the averaged nodal coordinates (centroid) of the $I^{th}$ cluster. This is essentially a discrete optimization problem with many possible local optimums whose final solution often depends on the initial guess (note that when applying the 'kmeans' function in MATLAB, we can specify the initial centroid values for each cluster to achieve the same clustering pattern in different trials). This clustering method can be applied to both macrostructures in Figure 2 and microstructures in Figure 3.

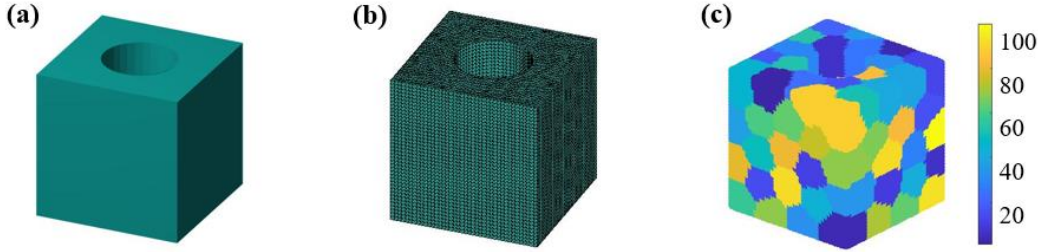

**Figure 3 Spatial domain decomposition converts a fine finite element (FE) mesh to a reduced cluster representation: (a)** An arbitrary structure. **(b)** The fine FE discretization. **(c)** The reduced representation is generated via the k-means clustering by agglomerating neighboring nodes as clusters. In this example, 100 clusters are generated where each cluster is a separate sub-domain indicated by the same color.

The number of clusters determines the data compression ratio, i.e., the reduction in DOF of the system. For example, the sufficiently fine mesh in Figure 3 has much more elements than clusters and hence its associated DOF is higher. Fewer clusters lead to a higher data compression level along with lower accuracy and computational costs. Hence, as we demonstrate in Section 5, one can start with a relatively small number of clusters and then increase this number until the predicted stresses converge (i.e., they don't change by further increasing the number of clusters).

We compare our clustering algorithm with the domain decomposition approach employed in the SCA method [16]. First, SCA groups material points with similar mechanical behaviors by applying orthogonal loading tests on microstructures where the pure normal or shear responses are computed. While such a decomposition works well for cubic microstructures, it may incur errors in complex macrostructures whose shapes are irregular. Our approach does not need *a priori* tests, i.e., it only relies on an existing mesh and is therefore applicable to any micro- and macro-structures. Second, the clustering algorithm in SCA allows topologically disconnected material points to belong to the same cluster by assuming points with similar elastic responses tend to behave similarly in plastic regimes. In our method, where a pre-computed mechanical response is unnecessary, we assume neighboring points behave alike plastically. Third, our clustering method is similar to SCA in that (1) material properties are assumed identical in the same cluster, and (2) clustering is performed separately for different material phases. In this paper, we investigate two-phase materials where the primary and secondary phases are the metal alloy and pores, respectively. Since pores do not involve material, domain decomposition is applied only to the metal alloy.

*3.2. Acceleration scheme of macroscale simulation*

Newton's method [43] is one of the most popular numerical techniques for iteratively solving the nonlinear problems in Equations (2) and (7). It successively updates the variables to better approximate the root(s) of nonlinear equations where a system of linear equations is solved in each iteration. Specifically, for a solid medium under static equilibrium conditions, the following linear system is solved:

$$\mathbf{K}\mathbf{u} = \mathbf{f} \qquad (27)$$

where **K** indicates the linear system's tangential stiffness matrix, **f** represents the unbalanced force between external and internal forces, and **u** is the incremental displacement solution. Since all quantities discussed in this section are macroscopic, the scale subscript 'M' is dropped for convenience. Since the linear system in Equation (27) is solved in each iteration of Newton's method, accelerating the solution process would significantly reduce the execution time of solving the underlying nonlinear problem.

To expedite the linear solution process, we first identify its major computational bottlenecks. A time comparison between different computational components is demonstrated in Figure 4 where the elastic response of the simple macrostructure in Figure 4(a) is simulated by classic FEM. It is observed from Figure 4(b) that the top three components that account for a significant portion (95.1%) of the total computational time include: solving a system of linear equations, computing elemental stiffness matrices, and assembling the global stiffness matrix. It is noted that using the FEM on a different geometry domain or using a distinct mesh size may change the absolute computational time of each component. However, it is found that the bottleneck of many simulations lies in the three components mentioned above [39]. Based on this observation, we improve the computational efficiency of these three components by adopting a rigid body cluster-based deflation method [44] combined with an incremental assembly technique [45]. These two methods are described in Section 3.2.1 and Section 3.2.2, respectively.

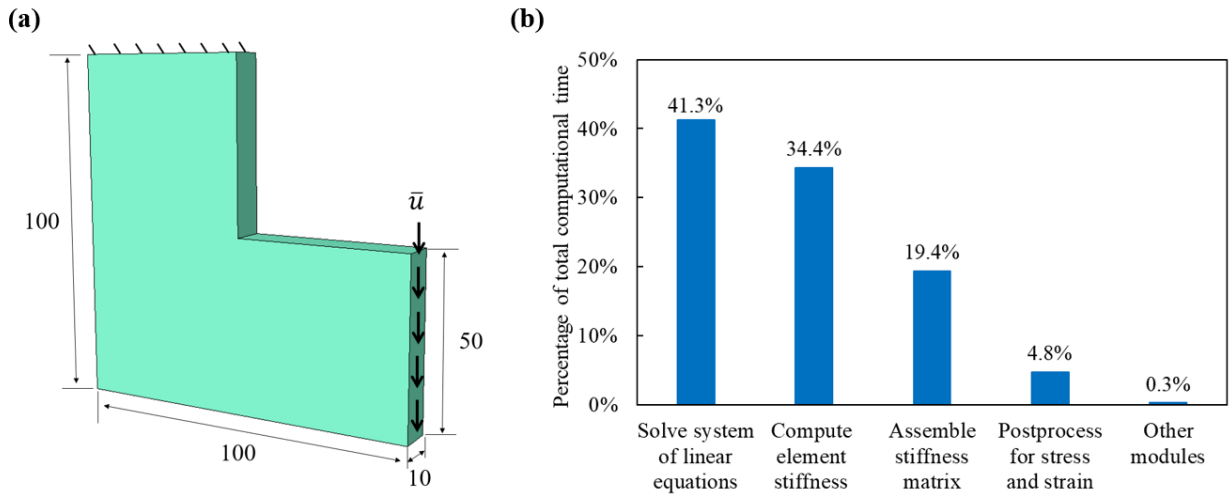

**Figure 4** Break down of the computational time of solving an incremental linear system by Newton's method: **(a)** A simple macrostructure (units: mm) is subject to a Dirichlet boundary condition ($\bar{u} = 1$ mm). **(b)** Different computational components correspond to different percentages of the total computational time.

*3.2.1. Clustering-based deflation method*

Conjugate gradient (CG) is often used in FEM to solve the algebraic system in Equation (27), especially when the number of DOF is large or the stiffness matrix dramatically changes across iterations where expensive matrix factorization cannot be reused. As CG is based on minimizing the energy norm of the system residuals in an iterative approximation over the Krylov space [46], its convergence rate depends on two factors: (1) the condition number of the system's stiffness matrix, and (2) the spectrum of its small eigenvalues. The two factors represent different natures of the studied problem. Two matrices can have the same condition number but the one with more small eigenvalues generally needs more CG iterations to converge. In a discretized solid continuum, the condition number generally increases as either the number of elements or the contrast between material properties increases. Hence, preconditioners are typically adopted to reduce the condition number of stiffness matrices. The diagonal of the stiffness matrix is a common choice for preconditioners since the associated computational costs and storage requirements are small.

While the eigenvectors associated with the smallest eigenvalues dominate the convergence rate to the global solution, it often requires a significant number of iterations to approximate them. In other words, the convergence of CG is generally slow for the low-energy modes associated with the small eigenvalues as they are insufficiently represented in the system residuals. To address this issue, we integrate our clustering technique with the rigid body-based deflation method (RBD) which was originally developed to expedite eigenvalue problems [47] and has also been used in computational solid mechanics [39]. The main idea behind the clustering-based RBD method is to construct a deflation matrix whose column vectors span the space of small (near-zero) eigenvectors that are approximated by the rigid body modes of clusters. Under the rigid body assumption, clusters are assumed to have zero strain energy and their displacements are in the null space of the stiffness matrix. The directions of rigid body modes are indicated by the basis vectors of the null space and their number equals the zero-value eigenvectors of the stiffness matrix. In essence, the deflation method projects the system's residual from the Krylov space to the deflation space, where the FE mesh's small eigenvectors are represented by the clusters' rigid body modes. 3D eigenmodes consist of three translations and three rotations. Since the small eigenvectors are readily available in the deflation space, the number of required CG iterations for convergence is significantly reduced.

To implement the RBD method, we first write the displacements of the clustering nodes as the rigid body motions of a cluster:

$$\mathbf{u_i^j} = \mathbf{W_i^j} \boldsymbol{\lambda_i} \tag{28}$$

where $\mathbf{u_i^j}$ is the displacement vector for the $i^{th}$ node in the $j^{th}$ cluster, $\boldsymbol{\lambda_i}$ represents the unknown vector of rigid body motions with six DOF in the $j^{th}$ cluster, and $\mathbf{W_i^j}$ is the deflation matrix defined for the $i^{th}$ node associated with the $j^{th}$ cluster. Specifically, $\boldsymbol{\lambda_i}$ and $\mathbf{W_i^j}$ are given as:

$$\boldsymbol{\lambda_j} = \left[ u_{jx}, u_{jy}, u_{jz}, \theta_{jx}, \theta_{jy}, \theta_{jz} \right]^T \tag{29}$$

$$\mathbf{W_i^j} = \begin{bmatrix} 1 & 0 & 0 & 0 & z_i^j & -y_i^j \\ 0 & 1 & 0 & -z_i^j & 0 & x_i^j \\ 0 & 0 & 1 & y_i^j & -x_i^j & 0 \end{bmatrix} \tag{30}$$

where $u_{jx}$ and $\theta_{jx}$ represent, respectively, the displacement and rotation of $j^{th}$ cluster centroid along the $x$ axis, and $(x_i^j, y_i^j, z_i^j)$ are the relative 3D coordinates of the $i^{th}$ node with respect to the $j^{th}$ cluster's centroid. The projection in Equation (28) works similarly to the restriction operation in the context of multigrid methods which map variables from a coarse mesh to a fine mesh.

It is noted the projection matrix $\mathbf{W_i^j}$ in the Equation (28) is defined for each node and assembly over all nodes is needed to construct the global deflation matrix $\mathbf{W}$ which projects the rigid body motions ($\boldsymbol{\lambda}$) to nodal displacements ($\mathbf{u}$) on the entire FE mesh as:

$$\mathbf{u} = \mathbf{W} \boldsymbol{\lambda} \tag{31}$$

If the number of nodes and clusters are $n_{nd}$ and $n_{cl}$, the dimensions of the vectors (**u** and **λ**) and the deflation matrix (**W**) in Equation (31) are $(3n_{nd} \times 1)$, $(6n_{cl} \times 1)$, and $(3n_{nd} \times 6n_{cl})$, respectively. By exploiting the global deflation matrix **W**, one can now implement the rigid body cluster-based deflated CG (DCG) as follows. In each Newton iteration, the linear system in Equation (27) is solved by splitting the displacement vector **u** into two parts [48]:

$$\mathbf{u} = (\mathbf{I} - \mathbf{A}^T)\mathbf{u} + \mathbf{A}^T\mathbf{u} \qquad (32)$$

where **A** is the projection matrix defined as [49]:

$$\mathbf{A} = \mathbf{I} - \mathbf{K}\mathbf{W}\mathbf{W}_d^{-1}\mathbf{W}^T = \mathbf{I} - \mathbf{K}\mathbf{W}(\mathbf{W}^T\mathbf{K}\mathbf{W})^{-1}\mathbf{W}^T \qquad (33)$$

The first part of Equation (32) can be extended using Equation (33) as:

$$(\mathbf{I} - \mathbf{A}^T)\mathbf{u} = \mathbf{W}\mathbf{W}_d^{-1}\mathbf{W}^T\mathbf{K}\mathbf{u} = \mathbf{W}\mathbf{W}_d^{-1}\mathbf{W}^T\mathbf{f} \qquad (34)$$

where $\mathbf{K_d} = \mathbf{W}^T\mathbf{K}\mathbf{W}$ is the deflated positive-definite stiffness matrix with dimensionality $(6n_{cl} \times 6n_{cl})$. It is projected from Krylov space onto deflation space, where the interaction components between FE nodes are condensed to the cluster-to-cluster interactions. In essence, since the number of clusters is much smaller than the number of nodes, i.e., $n_{cl} < n_{nd}$, the dimension of the deflated stiffness $\mathbf{K_d}$ is much smaller than its FE counterpart **K**. In such a scenario, matrix factorization and Gaussian elimination are directly applied at low costs to compute the inverse of the deflated matrix $\mathbf{K_d}$.

The second term of Equation (32) is computed by pre-multiplying both sides by $\mathbf{K}^T$:

$$\mathbf{K}^T\mathbf{A}^T\mathbf{u} = \mathbf{A}\mathbf{K}\mathbf{u} = \mathbf{A}\mathbf{f} \qquad (35)$$

where the symmetric property is utilized:

$$\mathbf{K}^T\mathbf{A}^T = \mathbf{A}\mathbf{K} \qquad (36)$$

In Equation (35), **AK** and **Af** are the projected stiffness matrix and force vector in the deflated space associated with the pre-defined rigid body modes. As discussed earlier, since the small eigenvectors are readily available in the deflation space, the number of CG iterations required to compute **u** in Equation (35) is considerably reduced. Once **u** is available, its projected counterpart ($\mathbf{A}^T\mathbf{u}$) can be easily computed and used as the second term in Equation (32). Therefore, by applying the RBD method, the displacement solution to the linear system in Equation (27) is readily available by combining the solutions from Equations (34) and (35).

In elastoplastic simulations, a sharp contrast between material properties in the presence of plastic yielding is a significant source for new eigenmodes which correspond to either large or small eigenvalues. While the small eigenvalues deteriorate the performances of other iterative solvers (e.g., CG), they do not affect the high efficiency of the DCG method we have adopted [50].

### 3.2.2. Incremental stiffness matrix assembly

Per Figure 4(b), computation of elemental stiffness matrices and assembly of the global stiffness matrix accounts for about 53.8% of the entire computational time. In elastoplastic simulations, the yielding changes material properties, and hence the material's stiffness matrix must be updated accordingly. Specifically, for 3D large-scale models, in each Newton iteration repeated allocation and deallocation of large computer memories are required for re-computing elemental stiffness matrices and re-assembling the global stiffness matrix. These intensive matrix operations increase memory footprints and slow down program execution.

We adopt the incremental assembly technique [45] to prevent the repeated matrix re-computing and re-assembly by only updating matrix entries associated with plastically yielded elements in each iteration of Newton's method. Our method is inspired by the observation that only a small portion of the material yields in many macroscale simulations involving plastic deformations. This behavior is because plasticity is generally localized at points with high strain concentrations which often result from geometric singularities or concentrated loads. For example, consider the same L-shape beam in Figure 4(a) subject to an elastoplastic deformation in Figure 5 (see Section 5 for material hardening behaviors). It is observed that the high values of Von-Mises stress and equivalent plastic strain are mainly located at the sharp corner (geometric singularity). In this example, the yielded elements only constitute 12.1% of the total number of elements.

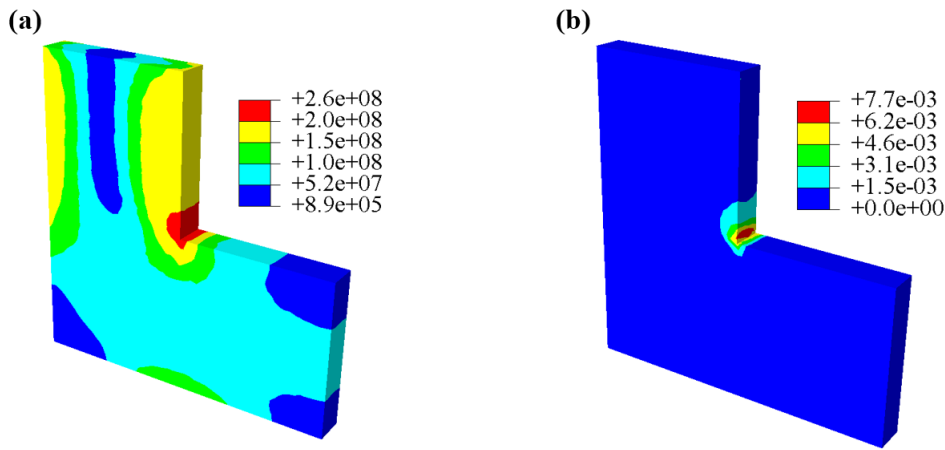

**Figure 5 Stress and strain fields of an elastoplastic analysis on a simple L-shape beam model: (a)** The distribution of Von-Mises stress where stress concentration locates around the sharp corner due to geometric singularity. **(b)** The distribution of equivalent plastic strain is highly correlated with stress-concentrated regions. In this model, only a fraction (12.1%) of elements is plastically yielded, so only their properties (as opposed to all elements) need to be updated in the global stiffness matrix.

Therefore, per Equation (37) we only compute and update a small number of entries in the global stiffness matrix instead of re-constructing the entire stiffness matrix in each Newton iteration:

$$\mathbf{K}^{\mathbf{i}} = \mathbf{K}^{\mathbf{i}\text{-}\mathbf{1}} + \Delta\mathbf{K}^{\mathbf{i}} \tag{37}$$

where $\mathbf{K}^{\mathbf{i}}$ and $\mathbf{K}^{\mathbf{i}-\mathbf{1}}$ are the global stiffness matrix at the $i^{th}$ and $(i-1)^{th}$ Newton iteration, respectively, and $\Delta\mathbf{K}^{\mathbf{i}}$ corresponds to the entries with updated material properties. As demonstrated in Section 5, our procedure significantly reduces the memory footprints and improves computational efficiency.

We combine the clustering-based deflation method with the incremental assembly technique to build the incremental deflated CG method (IDCG) which is summarized in Algorithm 1. It is noteworthy that applying clustering-based domain decomposition does not sacrifice the macroscale solution accuracy, as the same CG convergence criterion is enforced, see lines 26 and 36 of Algorithm 1. In essence, even though displacement fields are solved in deflation space for higher efficiency, they are projected back to Krylov space in each CG iteration for convergence check. In other words, the IDCG method generates the *exact* displacement solutions as the CG approach and as a result the deformation gradient at any macroscale integration point has high-fidelity and no solution accuracy loss.

**Algorithm 1** Structure of the macroscale rigid body cluster-based IDCG method

```
1:  procedure solve the macroscale linear system K^i u^i = f^i at the i^th Newton iteration
2:      ▷Initialization
3:      if i ← 1, then
4:          Generate FE mesh on the structure geometry
5:          Create node-based clusters via domain decomposition
6:          Initialize tangent stiffness matrix with elastic material properties
7:          Set incremental stiffness ΔK = 0
8:      else
9:          if plasticity occurs at the i^th Newton iteration, then
10:             Compute incremental stiffness ΔK
11:             K^i = K^{i-1} + ΔK^i
12:         else
13:             K^i = K^{i-1}
14:         end if
15:     end if
16:     Select a preconditioner M = diag(K^i)
17:     Choose an initial displacement u_0^i
18:     Set the initial residual r_0 = f^i − K^i u_0^i
19:     ▷Initialize state variables
20:     μ = (W^T K^i W)^{-1} W^T r_0
21:     u_1^i = u_0^i + Wμ
22:     r_1 = f^i − K^i u_1^i
23:     z_1 = M^{-1} r_1
24:     μ = (W^T K^i W)^{-1} W^T K^i z_1
25:     p_1 = z_1 − Wμ
26:     Set CG convergence criterion ε = 10^{-6}
27:     ▷Starting CG iterations
28:     for j← 1, N do     ▷Loop over N CG iterations
29:         α_j = (r_j z_j)/(p_j^T K^i p_j)
30:         u_{j+1}^i = u_j^i + α_j p_j
31:         r_{j+1} = r_j − α_j K^i p_j
32:         z_{j+1} = M^{-1} r_{j+1}
33:         β_j = (r_{j+1} z_{j+1})/(r_j z_j)
34:         μ = (W^T K^i W)^{-1} W^T K^i z_{j+1}
35:         p_{j+1} = z_{j+1} − β_j p_j − Wμ
36:         if ||r_{j+1}|| < ε, then
37:             ▷retrieve converged macroscale displacement solution at the i^th Newton iteration
38:             return u^i = u_j^i
39:         end if
40:     end for
41: end procedure
```

### 3.3. Acceleration scheme of microscale simulation

The incremental assembly technique discussed in Section 3.2.2 suits elastoplastic analyses of macrostructures well because a small portion of the material yields. However, in microscale simulations, a large number of elements may yield so we extend our RBD method to microscale as follows. We assume the strain field is uniform in each cluster and that it is computed based on the relative motions between interacting clusters. The uniform strain field assumption is similarly adopted by other ROMs. For example, TFA expresses the strain field as a linear combination of uniform eigenstrains in each material phase to reduce the number of state variables [12]; SCA computes the uniform cluster strains by the discrete Lippman-Schwinger equation which approximates the evolution of a cluster's strain field by considering its interactions with other clusters [16].

Our proposed method has three stages: building cluster-based computational mesh (Section 3.3.1), projecting solution variables between FE- and cluster-based mesh (Section 3.3.2), and constructing a reduced stiffness matrix for the cluster-based mesh (Section 3.3.3). Since all the quantities of interests discussed in this section are microscopic, the scale subscript 'm' is dropped.

*3.3.1. Construction of cluster-based mesh*

We build the computational grid by first considering cluster centroids as a set of scattering points and then connecting neighboring points via Delaunay triangulation (DT). By connecting four neighboring points we create tetrahedron elements that are based on clusters that agglomerate neighboring FE nodes. This way the topological features will be consistent between the FE- and cluster-based meshes. If one cluster is topologically connected to another, the motion of the first one should have a direct impact on the second one. In this scenario, we assume the two cluster centroids share the same DT element to account for their interactions. However, if two clusters are nearby but not topologically connected, the motion of one cluster should not have an immediate influence on the other one. In this case, we do not place the two cluster centroids in the same DT element.

Figure 6 illustrates the steps to generate the cluster-based mesh for the microstructure in Figure 3. In Figure 6(a) the microstructure is decomposed into 100 clusters whose connectivity is determined by checking if any of their nodes share the same FE. For example, it is observed in Figure 6(c) that cluster 1 is connected to cluster 2 as the FE nodes associated with the two clusters share the same FE on their boundary. On the contrary, it is evident that clusters 1 and 4 do not share any nodes belonging to the same FE and hence are not connected (even though the two clusters are close). This connectivity check between a cluster and its neighbors is examined for every cluster only *once* since the cluster-to-cluster connectivity relations are not changed during the analysis. We demonstrate cluster connectivity via a topological graph in Figure 6(d) where vertices and lines represent cluster centroids and topological connectivity, respectively. Based on the connectivity feature, we construct a computational grid in Figure 6(e) by connecting scattering cluster centroids in Figure 6(b) and ensuring all vertices (cluster centroids) in the same DT element are topologically connected per the topological graph of Figure 6(d).

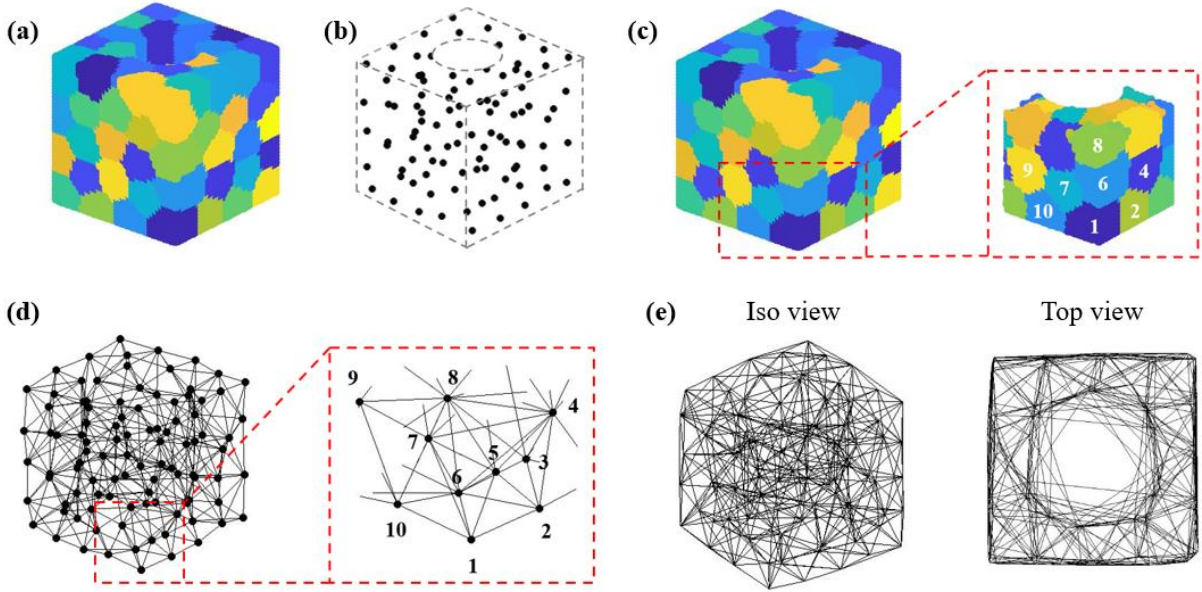

**Figure 6 Generation of cluster-based reduced mesh:** **(a)** Clusters are generated via domain decomposition. **(b)** The geometry centroids of the clusters. **(c)** An illustration of connectivity relations between clusters. **(d)** The connectivity relations are represented by a topological graph where each vertex represents a cluster centroid and lines indicate if two clusters are connected. **(e)** The reduced computational mesh is generated via DT by connecting topologically connected cluster centroids.

Unlike the SCA approach where cluster-to-cluster interactions are calculated using the Green function, we measure the interactions directly based on the clusters' position and topological connectivity. As we demonstrate in Section 5, by increasing the number of clusters our computational mesh converges to its FE counterpart. Specifically, in the limit when the number of clusters matches with the number of FE nodes, the reduced mesh resembles the FE mesh where each cluster only contains one node.

*3.3.2. Variable projections between FEM mesh and the cluster-based mesh*

In this section, we develop the mathematical formulations that project variables between the reduced mesh and its FE counterpart. Specifically, we define a restriction operator to map variables from FE mesh to the reduced mesh and a prolongation operator to project variables in the reverse direction.

As discussed previously, each cluster is a collection of neighboring nodes whose overall displacements are represented by the motion of the cluster's centroid. Therefore, given nodal displacements in a cluster, the restriction operator interpolates the nodal values onto its cluster centroid. While there are different types of interpolation methods such as polynomials [51] and Gaussian-based kernels [52], we adopt the polynomial augmented radial point interpolation method (PR-PIM) due to its simplicity, computational efficiency, and high accuracy. PR-PIM originates from the meshfree method [53] where its robustness is demonstrated by interpolating field variables over irregular scattering points.

Using PR-PIM, we approximate the displacement of the centroid of a cluster with its $n$ associated nodes via:

$$\mathbf{u}(\mathbf{x}) = \sum_{i=1}^{n} R_i(\mathbf{x}) a_i + \sum_{j=1}^{m} Z_j(\mathbf{x}) b_j \tag{38}$$

where $a_i$ is the coefficient of the radial basis function $R_i$ at the $i^{th}$ node and $b_j$ is the coefficient of the polynomial basis $Z_j$. The number of bases, $m$, is selected based on the polynomial reproduction requirement [53] which ensures Equation (38) can reproduce the solution of polynomial interpolation functions and hence helps to pass standard patch tests. For example, to meet a linear polynomial reproduction criterion in 3D, one only needs four polynomial bases, i.e., $m = 4$. When the number of nodes within a cluster is far more than the number of polynomial basis, i.e., $n \gg m$, the solution stability improves even if the nodes are irregularly distributed in the cluster [53].

We determine $a_i$ and $b_j$ coefficients by enforcing the interpolating function in Equation (38) to pass through all nodal values within a cluster. That is:

$$\mathbf{u}(\mathbf{x_k}) = \sum_{i=1}^{n} R_i(\mathbf{x_k}) a_i + \sum_{j=1}^{m} Z_j(\mathbf{x_k}) b_j, \quad k = 1, 2, \ldots, n \tag{39}$$

To ensure solution uniqueness [53], we force the polynomial terms to satisfy:

$$\sum_{i=1}^{n} Z_j(\mathbf{x}) a_i = 0, \quad j = 1, 2, \ldots, m \tag{40}$$

By combining Equations (39) and (40), we have:

$$\begin{bmatrix} \mathbf{R_Q} & \mathbf{Z_m} \\ \mathbf{Z_m^T} & \mathbf{0} \end{bmatrix} \begin{bmatrix} \mathbf{a} \\ \mathbf{b} \end{bmatrix} = \begin{bmatrix} \mathbf{d_s} \\ \mathbf{0} \end{bmatrix} \tag{41}$$

with

$$\mathbf{a} = [a_1, a_2, \ldots, a_n]^T, \quad \mathbf{b} = [b_1, b_2, \ldots, b_n]^T, \quad \mathbf{d_s} = [\mathbf{u}(\mathbf{x_1}), \mathbf{u}(\mathbf{x_2}), \ldots, \mathbf{u}(\mathbf{x_n})]^T \tag{42}$$

where $\mathbf{R_Q}$ is a symmetric moment matrix of the radial basis function (RBF) defined as:

$$\mathbf{R_Q} = \begin{bmatrix} R_1(r_1) & R_2(r_1) & \ldots & R_n(r_1) \\ R_1(r_2) & R_2(r_2) & \ldots & R_n(r_2) \\ \ldots & \ldots & \ldots & \ldots \\ R_1(r_n) & R_2(r_n) & \ldots & R_n(r_n) \end{bmatrix} \tag{43}$$

where $R_i(r_k)$ quantifies the radial basis function with distance $r_k$ between nodes $i$ and $k$. Additionally, $\mathbf{Z_m}$ is the polynomial-based moment matrix formulated as:

$$\mathbf{Z_m} = \begin{bmatrix} Z_1(\mathbf{x_1}) & Z_2(\mathbf{x_1}) & \ldots & Z_m(\mathbf{x_1}) \\ Z_1(\mathbf{x_2}) & Z_2(\mathbf{x_2}) & \ldots & Z_m(\mathbf{x_2}) \\ \ldots & \ldots & \ldots & \ldots \\ Z_1(\mathbf{x_n}) & Z_2(\mathbf{x_n}) & \ldots & Z_m(\mathbf{x_n}) \end{bmatrix} \tag{44}$$

The interpolation coefficients $\mathbf{a}$ and $\mathbf{b}$ in Equation (42) are solved as:

$$\mathbf{b} = \mathbf{S_b} \mathbf{d_s} \tag{45}$$

$$\mathbf{S_b} = \left[\mathbf{Z_m^T R_Q^{-1} Z_m}\right]^{-1} \mathbf{Z_m^T R_Q^{-1}} \tag{46}$$

and

$$\mathbf{a} = \mathbf{S_a d_s} \tag{47}$$

$$\mathbf{S_a} = \mathbf{R_Q^{-1}} - \mathbf{R_Q^{-1} Z_m S_b} \tag{48}$$

where $\mathbf{R_Q^{-1} Z_m}$ in Equation (48) is obtained by transposing $\mathbf{Z_m^T R_Q^{-1}}$. The interpolated displacement field of the cluster centroid is finally calculated by plugging Equations (45)-(48) back to the Equation (38):

$$\mathbf{u}(\mathbf{x}) = \left(\mathbf{R}(\mathbf{x})^T \mathbf{S_a} + \mathbf{Z}(\mathbf{x})^T \mathbf{S_b}\right) \mathbf{d_s} \tag{49}$$

The primary advantages of PR-PIM include: (1) its coefficient matrix in Equation (41) is always non-singular which guarantees the existence of a unique solution even when a cluster consists of a set of irregularly distributed nodes; (2) it preserves polynomial reproduction property which ensures its consistency with polynomial interpolation; and (3) its solution accuracy is not affected by the specific values of shape parameters in radial bases which removes the trial-and-error step associated with the parameter estimation of traditional RBF. Note that we perform the restriction operation only *once* at the beginning of microscale computations to map the nodal displacements onto cluster centroids. Detailed steps are included in Algorithm 2.

The prolongation operation maps the displacement field of a cluster's centroid back to finite element nodes. Since we represent nodal displacements with cluster centroid's rigid body motion, nodal displacement fields are computed by the deflation matrix in Equation (28). Similar to the restriction operation, the prolongation operation is executed only once after the microscale computations. As will be demonstrated in Section 5.2, the projected microstructural displacement fields resemble those obtained via FE as the number of clusters increases.

*3.3.3. Reduced stiffness matrix on the cluster-based mesh*

Once displacement variables are projected from the FE mesh to the cluster-based mesh, we construct the reduced stiffness matrix to account for the interactions between clusters. As demonstrated in Figure 6(e), the reduced stiffness matrix of the cluster-based mesh is constructed by creating reduced elements whose vertices represent neighboring clusters' centroids. FEM-based equilibrium equation is then applied to the reduced mesh to solve the displacements of the cluster centroids. In this manner similar to coarse-graining, the interaction components between FE nodes in different clusters are condensed to the interaction between vertices of a reduced element where mesh topologies are preserved between the FEM and cluster-based reduced meshes, see Figure 6.

On the reduced mesh, we assume there are six DOF at each element vertex, corresponding to the six 3D rigid body modes (three translations and three rotations). To account for rotations, we augment the stiffness matrix of classic tetrahedrons with rotational DOF. This augmentation can be achieved with a number of strategies. In [54], nodal rotations are introduced into four-node tetrahedrons by displacement-based local functions in a partition-of-unity approximation manner. In [55], Allmans rotational DOF along with a stabilization technique is introduced to control spurious element modes. In [56], the rotational DOF at the vertices of a four-node tetrahedron is approximated by transforming the mid-side translational DOF from a ten-node tetrahedron. A recent study [57] shows strain values in tetrahedrons can be significantly improved by adding virtual fiber rotations to regular displacement fields via the so-called space fiber rotation (SFR) concept. By comparing with classic solid elements, this study shows that the advantages of augmented tetrahedron elements are twofold: (1) their accuracy is globally close to that of classic quadratic elements but with a much higher computational efficiency; and (2) their accuracy is preserved quite well in coarse and distorted meshes. Since our reduced mesh is constructed from irregularly distributed points (cluster centroids), SFR-tetrahedron elements are ideal for our ROM.

As demonstrated in Figure 7, the SFR concept assumes each vertex of the four-node tetrahedron is associated with a virtual space fiber. The fiber rotation ($\boldsymbol{\theta}$) generates an additional displacement vector which enriches the classic element displacement field for an arbitrary internal point (q). The enriched displacement ($\mathbf{u_q}$) of the internal point (q) in a four-node SFR-tetrahedron is:

$$\mathbf{u_q} = \sum_{i=1}^{4} \mathbf{N_i} \left( \mathbf{u_i} + \boldsymbol{\theta_i} \otimes \mathbf{d_i} \right) \tag{50}$$

where $\mathbf{N_i}$ is the classic tetrahedron shape function on the $i^{th}$ node, $\mathbf{u_i}$ and $\boldsymbol{\theta_i}$ are, respectively, the displacement and rotation vectors on the $i^{th}$ node, and $\mathbf{d_i}$ is the relative position vector between the $i^{th}$ node and the internal point q. Following classic FE discretization [58], the stiffness matrix of the four-node SFR-tetrahedron is constructed as detailed in [57].

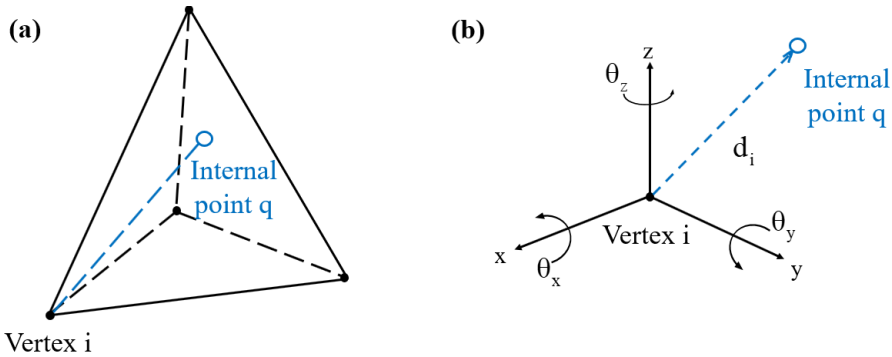

**Figure 7 Node and virtual space fiber representations:** (a) Schematic representation of the four-node SFR-tetrahedron. (b) Each SFR-tetrahedron vertex has three translational and three rotational DOF. The position vector ($d_i$) between the vertex (i) and the internal point (q) indicates the virtual space fiber.

The steps of the proposed microscale ROM are summarized in Algorithm 2 After the initial displacements are projected on the reduced mesh and the reduced stiffness matrix is constructed, we perform the microstructural elastoplastic analysis and compute stress and strain fields at each cluster centroid in the postprocessing step. Similar to TFA and SCA methods, where material points in one cluster have uniform stress and strain values, we assume all agglomerated nodes share the same stress and strain values as their cluster centroid. In other words, instead of computing distinct stress and strain fields at different elements, material points nearby (in one cluster) are assumed to share the same states (stress and strain). Hence, compared to classic FEM, the number of unknown variables in our ROM is significantly smaller, which dramatically improves computational efficiency while preserving effective accuracy. We point out that there is a major difference between the microscale analyses in Algorithm 2 with the macroscale solution process in Algorithm 1. To serve the purpose of multiscale simulations, the macroscale clustering-based deflation method aims to compute the exact local deformations at each macro-integration point, while microscopic projection analysis targets to obtain accurate homogenizations by coarse-graining microscale local responses. With high-fidelity macroscopic deformation gradients and effective responses, the accuracy of multiscale modeling is guaranteed.

**Algorithm 2** Structure of the microscale cluster-based reduced-order modeling

```
1:  procedure solve for the homogenized microstructural responses at the i^th nonlinear increment
2:      ▷Initialization
3:      if i ← 1, then
4:          Generate FE mesh on the microstructure
5:          Compute the microstructure geometric center x_center from nodal coordinates x
6:          Create node-based clusters via domain decomposition (Section 3.1)
7:          Construct cluster-based reduced mesh (Section 3.3.1)
8:          Initialize material properties with elastic material properties
9:          Develop reduced stiffness matrix on cluster-based mesh (Section 3.3.3)
10:     else
11:         Update material properties from the last increment
12:         Update the cluster-based constitutive model
13:     end if
14:     Read macroscopic deformation gradient F^i(X)
15:     Compute incremental homogeneous displacements on FE mesh: u_0^i(x) = (F^i(X) − I)(x − x_0)
16:     Project the homogeneous displacements from FE mesh to cluster-based mesh (Section 3.3.2)
17:     ▷Starting Newton iterations to solve micro-equilibrium equations on the reduced mesh
18:     Set Newton iteration convergence criteria ϵ = 10^{−3}
19:     for j ← 0, N do    ▷Loop over N Newton iterations
20:         Compute the microscale internal force vector f_j^{int}(u_j^i(x))
21:         Solve micro-equilibrium for microscopic displacement fluctuation ũ_j(x) (Section 2.2)
22:         Update microscale displacements: u_{j+1}^i(x) = u_j^i(x) + ũ_j(x)
23:         Postprocess for microstructural stress and strain fields
24:         ▷check for convergence
25:         if iteration residual < ϵ, then
26:             Compute homogenized stress tensor and material modulus (Section 2.3 and 2.4)
28:             return
29:         end if
30:     end for
40: end procedure
```

## 4. Microstructure characterization and reconstruction

Porosity is a common process-induced defect that significantly affects material behavior in cast metallic alloys. Since local morphological details of pores often randomly vary at the macroscale (see Figure 1), we develop a stochastic microstructure characterization and reconstruction (MCR) technique to investigate the effect of microscopic pores and their spatial distribution on the macroscopic response of materials. In this context, microstructure characterization involves building a statistical representation for the heterogeneous pore morphologies which is subsequently used in the reconstruction process to generate microstructures whose randomly distributed pores follow a desired distribution [59]. We adopt a descriptor-based MCR approach that characterizes morphological randomness via a carefully selected small set of physical descriptors defined in the characterization stage. Such an approach is very advantageous in building process-structure-property links in many material systems such as alloys [60].

Physical descriptors are either deterministic or statistical. A deterministic descriptor often only requires a single value to characterize the entire microstructure (such as pore volume fraction) while a statistical descriptor uses a distribution to characterize the spatial randomness of a morphological feature (e.g., the distribution of pore sizes). The values of physical descriptors are either estimated from microstructure images or selected via the design of experiments (DOE). In the first approach, image segmentation techniques are first applied to detect a set of pre-determined morphological features. The features are then analyzed to calculate the specific values of deterministic descriptors or the distribution parameters of statistical descriptors. If the above two steps result in too many descriptors, the most important ones can be identified through dimension reduction techniques [10]. The DOE-based approach is typically adopted in computational studies and building generalized process-structure-property maps [64,65]. In this approach, which is adopted in this work, a set of descriptors that sufficiently characterize the morphological features are first selected based on domain knowledge and application. Then, sample descriptor values are generated via DOE while considering their feasibility and practicality (e.g., a microstructure with a solid cluster that is topologically disconnected from its surrounding is not physically feasible). Finally, microstructures corresponding to each DOE point are reconstructed and used in simulations.

We model pore shapes as prolate ellipsoids with two identical minor axes and use four physical descriptors to describe the shape and spatial distribution of the pores in an RVE: pore volume fractions ($V_f$), number of pores ($N_p$), aspect ratio between major and minor axes ($A_r$), and the average spatial distance between two nearest pores ($\bar{r}_d$, units in μm). In our studies, these descriptors sufficiently characterize the effect of morphology on the homogenized response of porous microstructures. We set the ranges of these descriptors for DOE as:

$$V_f = 6.5\%, \quad N_p = [5, 100], \quad A_r = [1, 5], \quad \bar{r}_d = [10, 30] \tag{51}$$

We use the Sobol sequence [63] to sample from the ranges in Equation (51) because it very efficiently builds space-filling designs whose projections on any hyperplane are guaranteed not to overlap. We also make the following assumptions when reconstructing the virtual microstructure corresponding to a DOE point: (1) each microstructure is periodic and has a side length of 100 μm; (2) pores can overlap and are assumed to have similar sizes and shapes; (3) pores are assumed randomly dispersed and oriented; and (4) the lengths of prolate ellipsoid axes (major axis $r_a$ and minor axis $r_b$, units in μm) are smaller than the half of the microstructure size and large enough to avoid excessive nonlinearities:

$$r_a \geq 1.1, \quad r_b \leq 50 \tag{52}$$

where half of the microstructure size is assumed as 50 μm. Once the descriptors are selected and their values are determined via DOE, we reconstruct the microstructure corresponding to each DOE point via an optimization process that iteratively adjusts an initial microstructure until its descriptors match with the DOE point. As demonstrated in Figure 8, our reconstruction method has a hierarchical nature [64] and starts by assigning the deterministic high-level descriptor, i.e., the number of pores. It then adjusts pore locations to obtain the desired averaged distance between the nearest neighbors (i.e., $\bar{r}_d$). The adjustment of pore-to-pore distance is often achieved via a heuristic optimization algorithm such as simulated annealing. If a pore intersects with a microstructure boundary, an identical pore is added on the opposite side of the boundary to enforce periodicity. In the next level of reconstruction, geometric features, such as orientation and aspect ratio are assigned to each pore. Finally, pore sizes are proportionally calibrated to compensate for the overlaps and meet pore volume fraction requirements. Compared to the previous work [64] which focuses on 2D composites (polymer matrix with carbon fibers), our work extends the method to 3D and uses it to generate porous microstructure. This extension requires addition of more descriptors (to characterize 3D topologies instead of 2D) as well as detection and removal of isolated solid regions (because a floating solid portion is not physically meaningful). Six sample reconstructed microstructures are demonstrated in Figure 9 which indicates that our MCR method is capable of building microstructures with a wide range of porosity distribution.

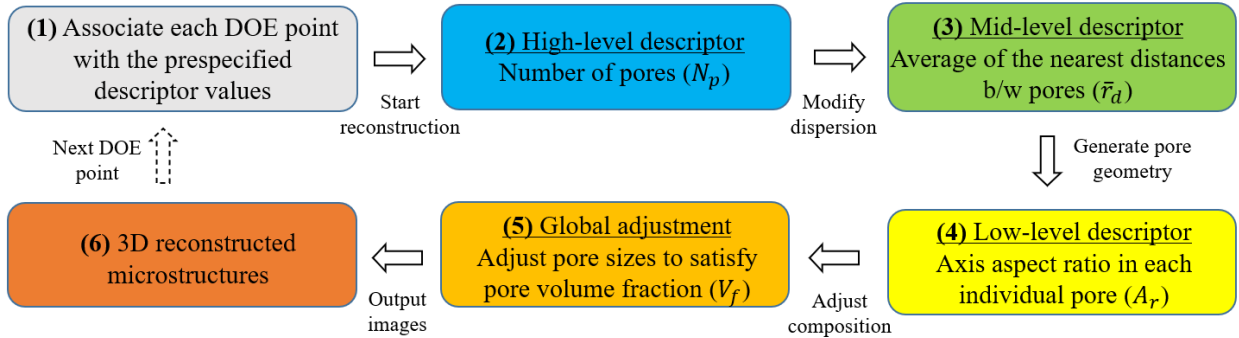

**Figure 8 Our MCR flowchart:** We develop a hierarchical microstructure reconstruction algorithm based on pore physical descriptors.

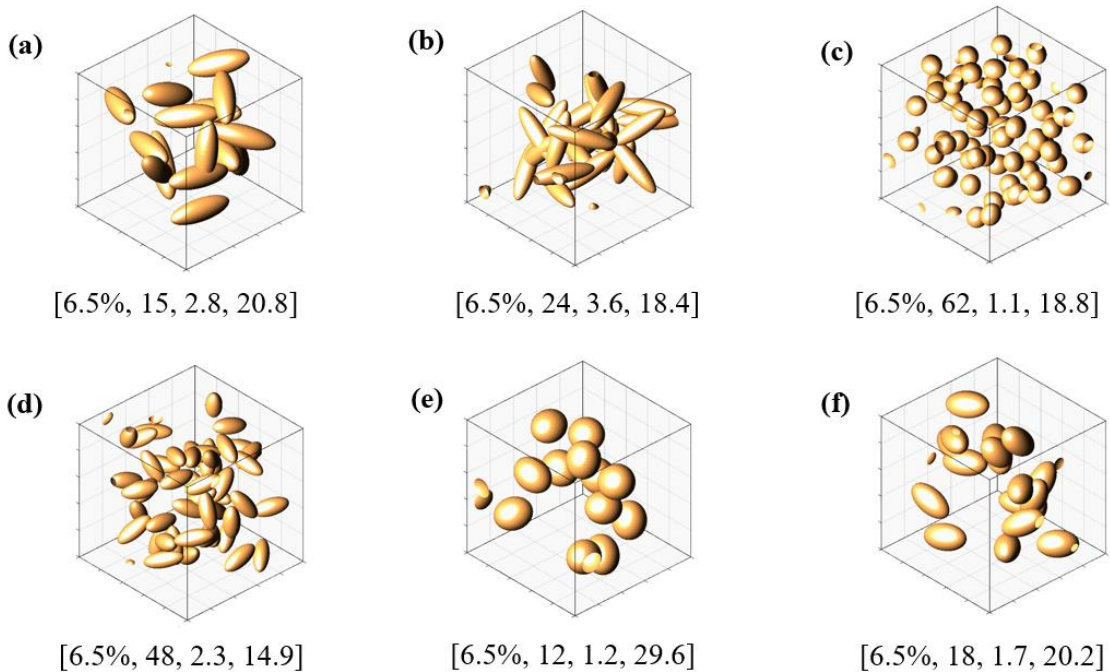

**Figure 9 Sample microstructures:** Six periodic microstructures are reconstructed per Figure 8 where the ranges of descriptors are defined in Equation (51). The primary material phase is not shown for clarity and a vector of the physical descriptor values $[V_f, N_p, A_r, \bar{r}_d]$ are listed under each microstructure.

## 5. Numerical experiments

In this section, we use the proposed DCA method to study the effect of porosity on the nonlinear elastoplastic behaviors of manufactured metallic components made out of aluminum alloy A360, which is a type of die casting alloy with excellent pressure toughness and high strength even in elevated temperatures. We consider A360 as the primary phase and the porosity as the secondary phase which is assumed to be the only microstructural defect. Spatial domain decomposition introduced in Section 3.1 is only performed on the primary phase in all micro-, macro-, and multi-scale simulations. Considering other metal polycrystalline microscopic features such as grain boundaries are out of the scope of this work.

The values of elastic modulus ($E$) and Poisson's ratio ($v$) are given as:

$$E = 6.89E4 \text{ MPa}, \quad v = 0.35 \qquad (53)$$

The elastoplastic behavior of A360 is assumed to follow the Von-Mises yield surface as:

$$\bar{\sigma} \leq \sigma_Y(\bar{\varepsilon}) \qquad (54)$$

where $\bar{\sigma}$ is the Von-Mises equivalent stress and the yield stress $\sigma_Y$ is governed by a predefined hardening law that depends on the equivalent plastic strain $\bar{\varepsilon}$. The material hardening behavior of A360 integrated into our simulations is demonstrated in Figure 10.

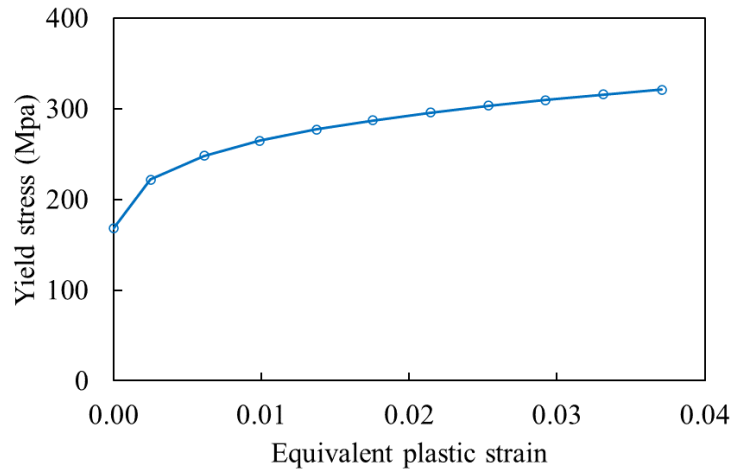

**Figure 10 Hardening behavior of A360:** A piecewise linear hardening is used in our simulations.

As discussed in Section 3, our DCA framework comprises two major components that accelerate macroscale and microscale simulations. Correspondingly, in Section 5.1, we use a macroscopic 3D bracket model to demonstrate the benefits of the clustering-based incremental deflation method discussed in Section 3.2. Next, in Section 5.2, we use our MCR algorithm in Section 4 to generate various 3D microstructures that embody a wide range of porosity characteristics. We then use our microscopic projection method introduced in Section 3.3 to deform these reconstructed samples following complex load paths. Finally, in Section 5.3, we combine the above two acceleration schemes within the first-order computational homogenization framework to quantify the effects of spatially distributed micro-porosity on macroscopic component behavior. The accuracy of the proposed ROM is verified against direct numerical simulations (DNS).

The proposed method is implemented in MATLAB, and all experiments are conducted on a 64-bit Windows workstation with the following hardware: Intel E5-2643 CPU with 12 cores running at 3.5 GHz with 128 GB installed physical memory (RAM) and 128 GB virtual memory.

*5.1. Macroscale experiments*

We test the rigid body cluster-based incremental deflation (IDCG) method on a macroscale 3D bracket shown in Figure 11(a), where one of its ends is fixed and two Dirichlet boundary conditions ($\bar{u} = 2\ mm$) are applied on the tips of the other end. We mesh the bracket with 180,000 finite elements to accurately find the distribution of Von-Mises stress upon elastoplastic deformation, see Figure 11(b). In this simulation, 5.1% of the elements yield, which are mainly located around the hole.

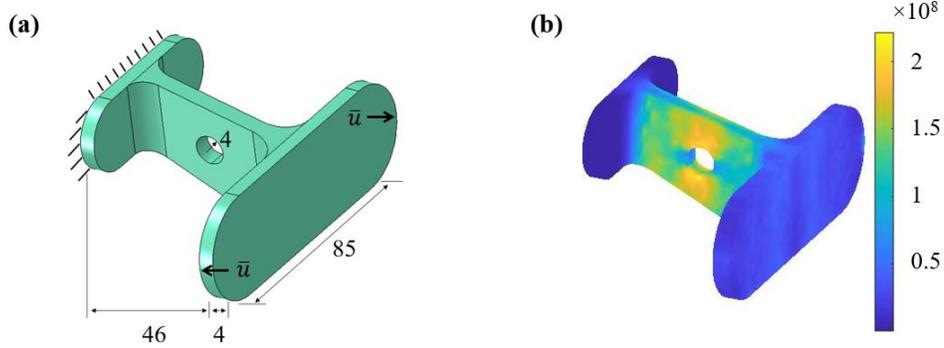

**Figure 11 Macroscale model: (a)** Geometry, dimension (unit: mm), and boundary conditions of the 3D bracket. **(b)** The distribution of Von-Mises stress (unit: Pa) after the elastoplastic simulation.

To assess the effectiveness of our deflation method, the bracket model is decomposed into three clustering models with $k = 50, 100, 200$ clusters, respectively, see Figure 12, which are then subject to elastoplastic simulations.

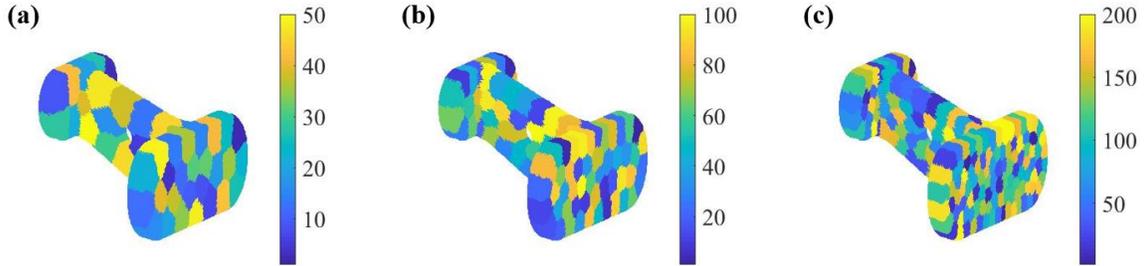

**Figure 12 Domain decomposition on the bracket model: (a)** 50 clusters. **(b)** 100 clusters. **(c)** 200 clusters.

We first demonstrate the efficiency of the deflation method (DCG) in reducing CG iterations by analyzing one Newton incremental solution process. The basic idea of using CG to solve an algebraic system, e.g., in Equation (27), is to iteratively minimize the vector difference between its left and right sides, or the residual as:

$$\mathbf{r}_{j+1} = \mathbf{r}_j - \alpha_j \mathbf{K}^i \mathbf{p}_j \tag{55}$$

where $\mathbf{r}_{j+1}$ is the CG residual at the current $(j+1)^{th}$ CG iteration, $\mathbf{r}_j$ is the error at the last $j^{th}$ iteration, $\mathbf{K}^i$ is the tangent stiffness at the $i^{th}$ Newton iteration, $\alpha_j$ and $\mathbf{p}_j$ are, respectively, the iterative size and vector at the $j^{th}$ CG iteration, see Algorithm 1. By enforcing the same CG convergence criterion ($\|\mathbf{r}_{j+1}\| \leq \epsilon = 10^{-6}$) in all calculations, we guarantee the displacement solutions from DNS (CG) are recovered by the deflation methods and so are the stress and strain fields via postprocessing. With high-fidelity displacement results, accurate local deformation gradients are computed at each macroscale integration point which are subsequently passed to microstructural analyses in Section 5.2.

The efficiencies of DNS and deflation methods are compared in Figure 13(a). We observe that more than 2000 CG iterations are required for convergence by DNS. By comparison, our deflation method needs less than 100 CG iterations to achieve the same convergence criterion ($\epsilon = 10^{-6}$), showing a reduction of CG iterations by 20 folds. Specifically, as $k$ increases from 50 to 200, the required CG iterations drop from 95 to 52, indicating that the CG residual is more efficiently reduced in the deflation space constructed by 200 rigid bodies. It is also evident that DNS shows many stagnation stages in Figure 13(a), e.g., from iterations 500 to 1000, which is due to the existence of multiple near-zero eigenvalues whose approximation is difficult, see Figure 13(b). These stagnation stages are not observed in the deflation method since it projects CG residuals into the deflation space where the system's small eigenvalues are readily represented by the clusters' rigid body motions. In addition, we observe the clustering systems are better conditioned with smaller condition numbers, measured as the ratio between the maximum and minimum eigen values, than that of the DNS from Figure 13(b). Smaller condition numbers, no near-zero eigenvalues, and fewer degrees of freedom together explain the higher efficiency of the clustering-based deflation approach.

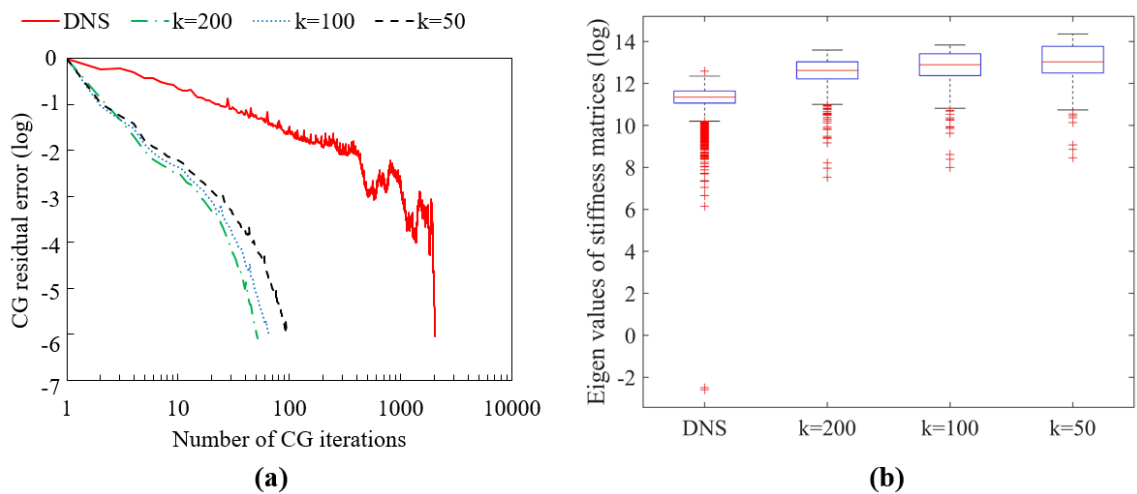

**Figure 13 Macroscale solver comparisons: (a)** CG Convergence of DNS is compared against the deflation method (DCG) with different cluster numbers (k) in one Newton incremental solution process; **(b)** Distributions of eigen values of the underlying stiffness matrices by box plots where red crosses demonstrate the extreme values of eigen values where the number of eigen values of DNS, k=200, 100, and 50 are 12942, 1200, 600, and 300, respectively. The near-zero eigen values in DNS are removed in the clustering-based deflation methods.

We now assess the efficiency of the incremental assembly technique by recording the total time for computing element stiffness matrices and assembling the global stiffness matrix in an elastoplastic simulation. To this end, we use the bracket model in Figure 11(a) and mesh it with different numbers of elements. Figure 14(a) compares the costs associated with our approach with the traditional full-size assembly. It is observed that the classic stiffness computation and assembly approach becomes rapidly expensive as the number of elements increases. However, with our method, these costs scale much slower since we only update stiffness entries associated with the yielded elements, which only account for 5.1% of the entire structure on the mesh with 180,000 finite elements. We note that the time reported in Figure 14(a) does not depend on the number of clusters since stiffness matrix computation and assembly are conducted before online solutions where deploying clusters shows significant acceleration.

Our macroscale ROM (IDCG) benefits from both acceleration schemes discussed above. To quantify its overall efficiency, we compare the total computational time of our IDCG method against DNS in Figure 14(b) for elastoplastic simulations. The comparison is performed on different mesh sizes while ensuring each simulation achieves the same CG residual as in Figure 13(a). As demonstrated, our method is not only faster than DNS on all mesh sizes but also scales more favorably, i.e., as the number of elements increases, the computational savings of our approach grow. We note that cost savings in a macroscale simulation are not directly related to the number of clusters used in the IDCG method. This behavior is because increasing cluster numbers results in fewer CG iterations, but a larger size of deflation system whose sparse matrix-vector multiplication (SpMV) [46,52] is computationally more expensive.

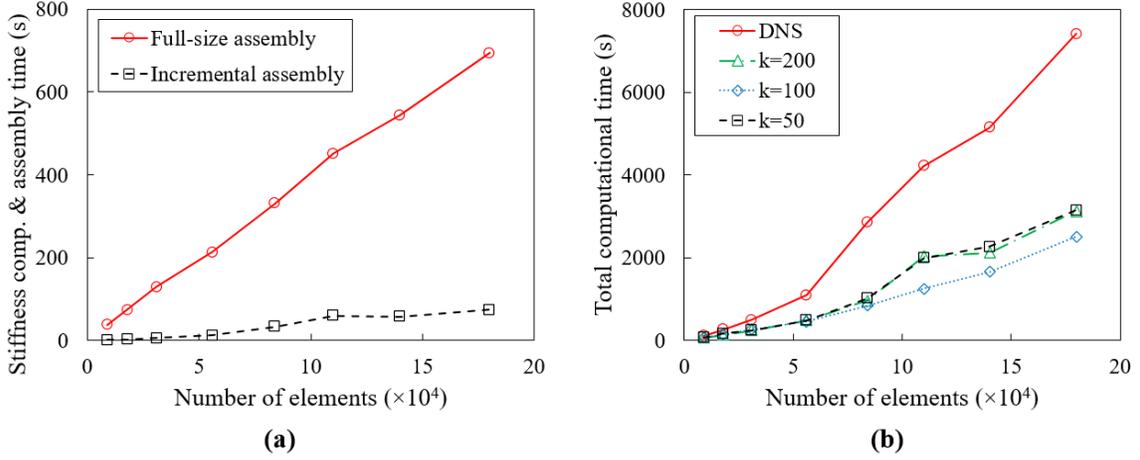

**Figure 14 Efficiency comparisons:** (a) Time reduction on stiffness matrix computation and assembly via the incremental assembly technique, and (b) Computational time of classic FEM (DNS) using pure CG against the incremental assembly deflated CG method (IDCG) with different cluster numbers (k).

*5.2. Microscale experiments*

We test the performance of our ROM on multiple microstructures subject to complex loading paths that include high plastic deformations and cyclic loading. The first test case is a microstructure that has a single-cylinder hole in the center corresponding to a pore volume fraction of 19.6%, as shown in Figure 15. Because this microstructure has a simple geometry, we utilize it as a benchmark to demonstrate the property of the reduced mesh. We assume the microstructure is subject to a multiaxial deformation as in Equation (56). The resultant equivalent plastic strain fields are compared with the FEM results as in Figure 15.

$$\mathbf{F} = \begin{bmatrix} 1.02 & 0 & 0 \\ 0 & 0.99 & 0 \\ 0 & 0 & 0.99 \end{bmatrix} \quad (56)$$

As illustrated in Figure 15, we decompose the microstructural domain into different numbers of clusters. The reduced mesh gradually converges to its FE counterpart and closely represents the microstructural geometry with the increase of cluster numbers. This mesh consistency is because the cluster connectivity relations are well preserved between the FE- and reduced- meshes.

Along with the computational meshes, we also compare the distributions of equivalent plastic strain fields computed on the reduced meshes against their FE counterpart. It is observed that as the number of clusters increases, the plastic fields obtained by our reduced-order model converge to that obtained via the FE method. This convergence is on average, i.e., the local values are slightly different as similarly reported in [16]. In our case, the local differences can be explained by the fact that we assign the same strain value to each cluster while FEM is free to produce significant strain gradients in the small regions that experience high concentrations. Averaging this local information is in essence a lossy information compression that endows our method with lower computational costs and memory footprint compared to FEM. To improve local prediction accuracy, a straightforward approach is to increase the average number of clusters as suggested in [19,24]. However, a better strategy is to only increase clusters in the regions with high stress or strain concentrations. Such regions can be detected by elastic analysis in a preprocessing stage or on the fly. We will pursue this direction in our future works.

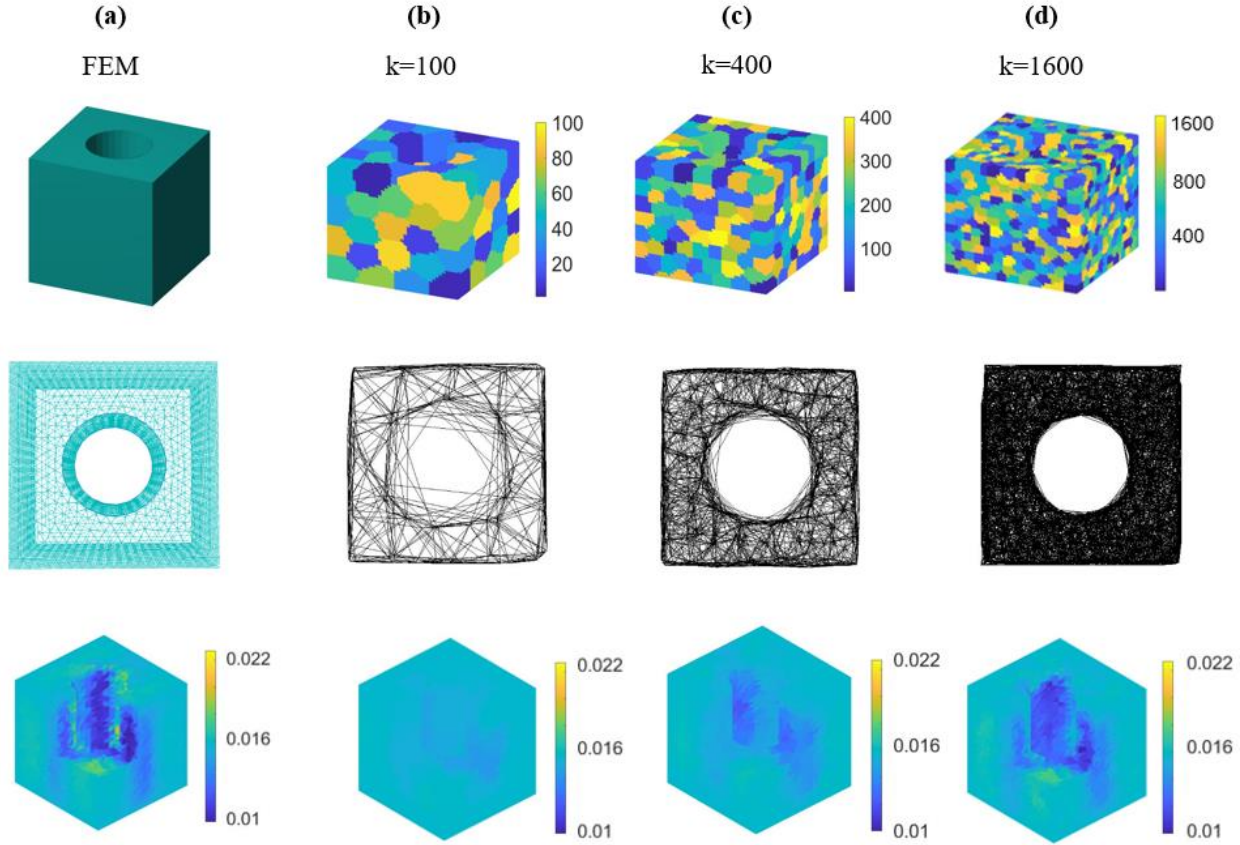

**Figure 15 Influence of cluster numbers on the accuracy of equivalent plastic strain field:** Column **(a)** shows the microstructure domain, its FEM mesh (top view) with 40,482 elements, and its plastic strain fields in sequence. Column **(b)** sequentially shows the domain decomposition with 100 clusters, the cluster-based reduced mesh (top view), and the distribution of equivalent plastic strain. Columns **(c)** and **(d)** correspond to 400 and 1600 clusters, respectively.

We now test the proposed ROM quantitatively on reconstructed microstructures with complex pore morphologies, see Figure 16. The first experiment tests whether the proposed model can accurately predict microstructural homogenized nonlinear responses when the microstructure is subject to the complex deformation state given in Equation (57). The studied microstructure is shown in Figure 16(a) and has 74 pores that possess a volume fraction of 7.6%. For DNS, the microstructure is discretized by 421,507 linear tetrahedrons with 232,692 DOF. With the microstructural domain decomposed into different numbers of clusters, we compare the reduced model's homogenized responses with DNS.

$$\mathbf{F} = \begin{bmatrix} 1.01 & 0.02 & 0.025 \\ 0.02 & 1.02 & 0.03 \\ 0.025 & 0.03 & 0.97 \end{bmatrix} \tag{57}$$

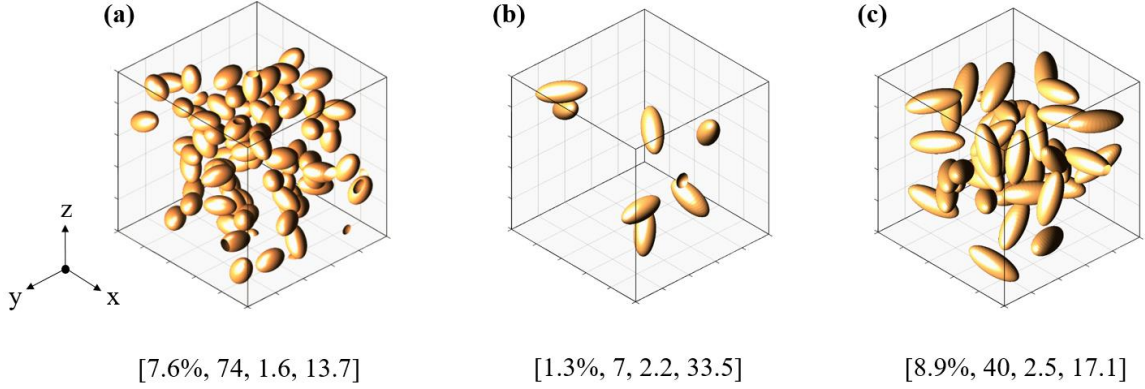

[7.6%, 74, 1.6, 13.7]     [1.3%, 7, 2.2, 33.5]     [8.9%, 40, 2.5, 17.1]

**Figure 16 Reconstructed microstructures:** Morphologically different samples whose specific descriptor values $[V_f, N_p, A_r, \bar{r}_d]$ are listed in a vector below the image.

The predicted homogenized stress components are compared with DNS results in Figure 17. It is observed that the homogenized stress components from the proposed ROM gradually converge to the DNS results as the number of clusters increases. The displacement fields illustrated in Figure 17 depict the same behavior where with few clusters there is a large discrepancy between the nodal displacements close to the cluster boundaries. As the number of clusters increases, the discrepancy between the displacement fields across cluster boundaries diminishes and the global domain resembles the DNS results.

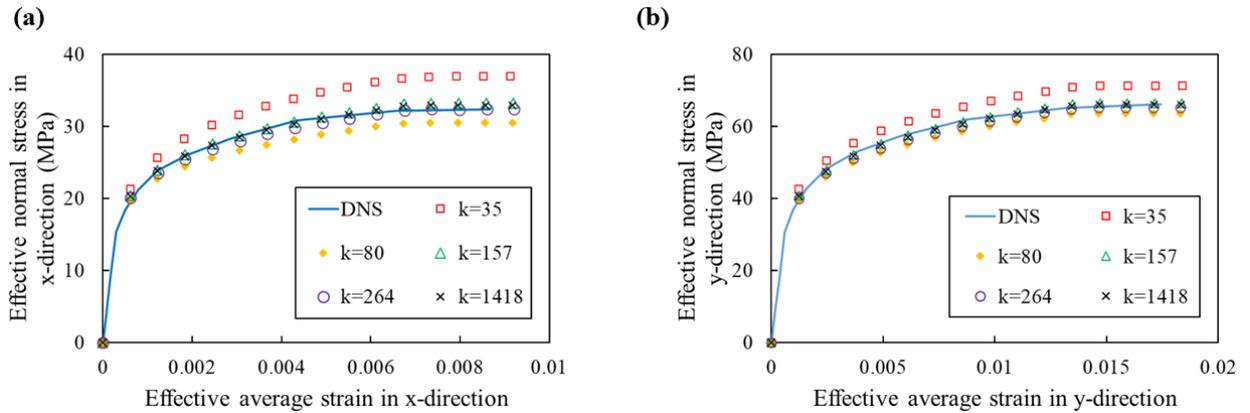

**Figure 17 Homogenized stress components for the microstructure in Figure 16(a): (a)** The normal stress component in the x-direction. **(b)** The normal stress component in the y-direction. The stressed are obtained via DNS and our ROM with different cluster numbers.

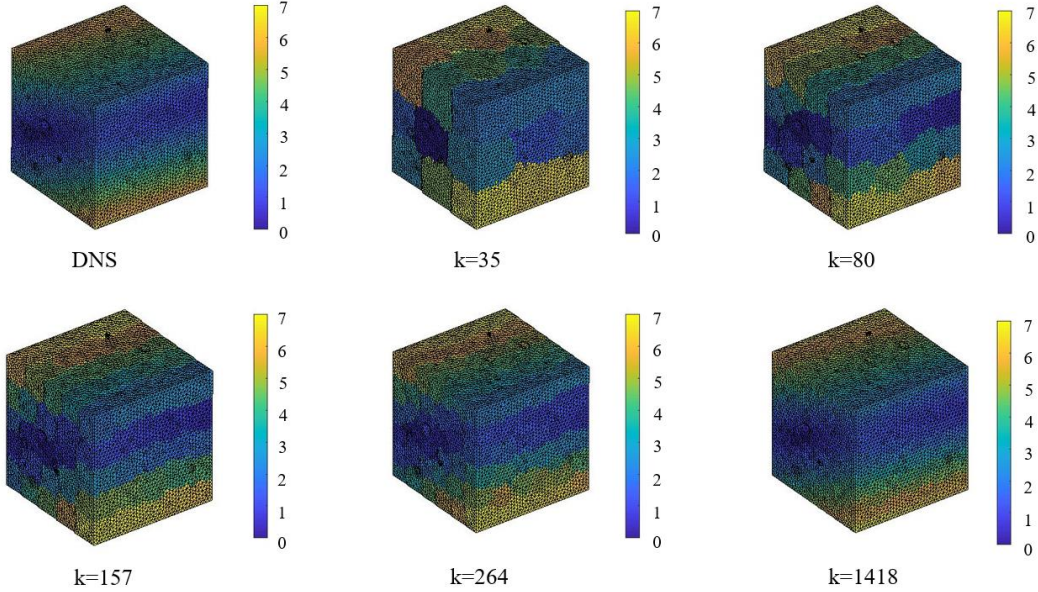

**Figure 18 Comparison of displacement fields (unit: μm):** The fields are for the microstructure in Figure 16(a) between DNS and our ROM with different numbers of clusters.

We compare the computational costs of our approach and DNS for this example in Figure 19. As it can be observed, with 264 and 1418 clusters we accelerate the simulations by more than 55 and 10 times, respectively. It is worth noting that while the DNS is performed on a highly optimized commercial software package (ABAQUS [65]), our method is implemented in MATLAB scripts and can greatly benefit from optimizing memory footprints or utilizing high-performance computing techniques [21]. Also observed in Figure 19 is the strong dependence of the computational time on the number of clusters which indicates that updating cluster-wise state variables accounts for most of the costs in the ROM.

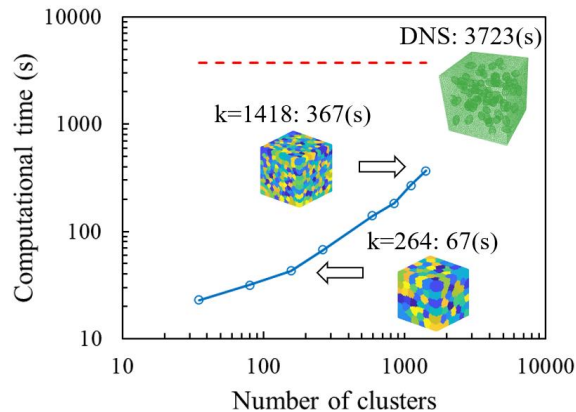

**Figure 19 Effect of cluster number on costs:** Comparison is between DNS and our ROM with different clusters for the microstructure in Figure 16(a).

In the second experiment, we test our model on a complex loading path on the microstructure shown in Figure 16(b) which has seven pores and a pore volume fraction of 1.3%. In DNS, the microstructure is discretized by 234,573 linear tetrahedrons that result in 129,291 DOF. We assume this microstructure starts from a relaxing initial state and is then subject to the two-step deformation gradient constructed with $\mathbf{F}_1$ and $\mathbf{F}_2$:

$$\mathbf{F}_1 = \begin{bmatrix} 1.01 & 0.005 & 0.01 \\ 0.005 & 1.02 & 0.015 \\ 0.01 & 0.015 & 0.97 \end{bmatrix}, \quad \mathbf{F}_2 = \begin{bmatrix} 0.97 & 0.015 & 0.02 \\ 0.015 & 1.01 & 0.005 \\ 0.02 & 0.005 & 1.02 \end{bmatrix} \tag{58}$$

The effective stress-strain relations are illustrated in Figure 20. Similar to the previous experiments, we observe that as the number of clusters increases the error with respect to DNS decreases. In particular, sufficiently accurate results are obtained with 264 clusters.

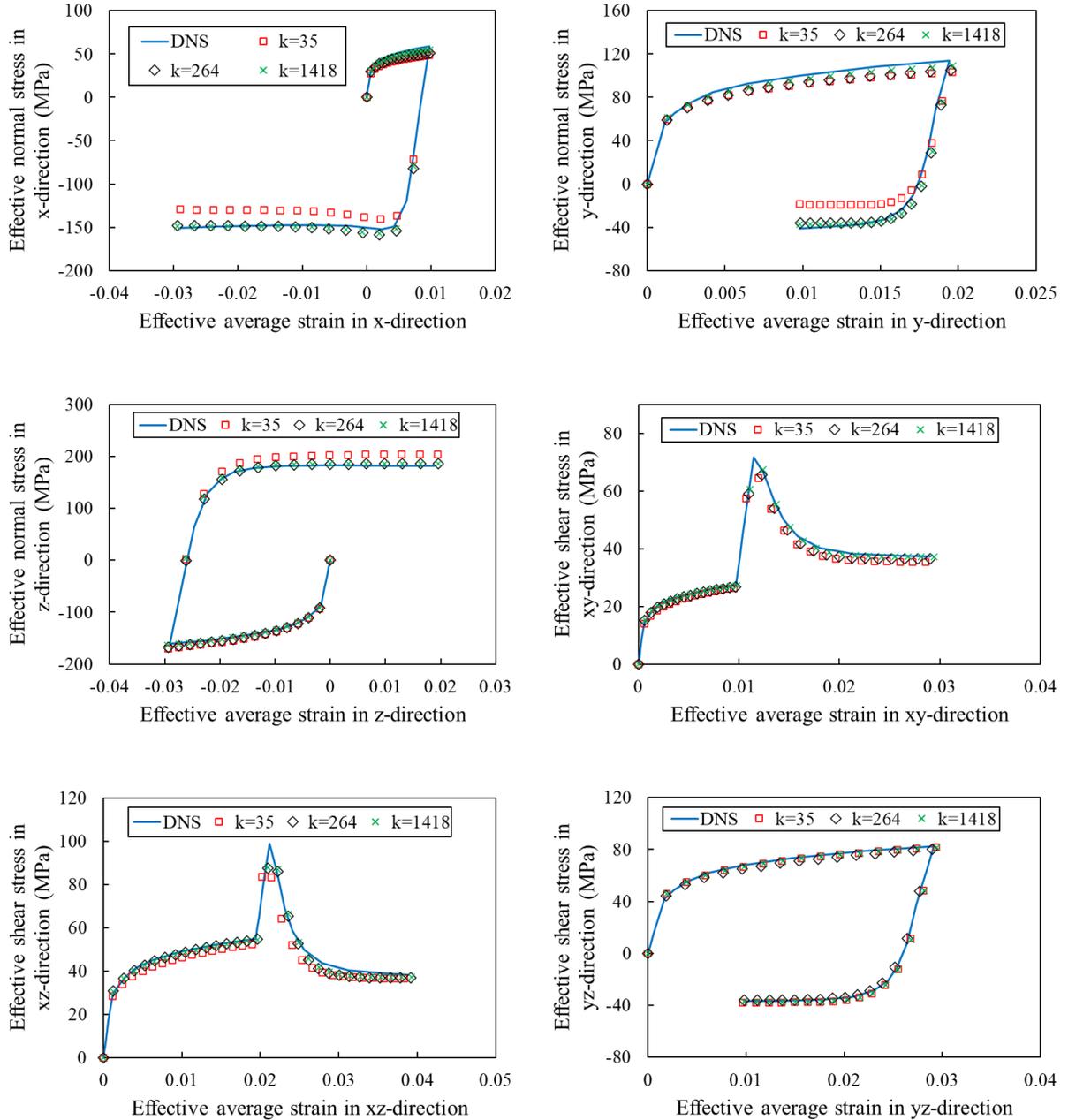

**Figure 20 Comparison of the homogenized stress components:** The comparison is between DNS and our ROM with different numbers of clusters on a two-step loading path for the microstructure in Figure 16(b).

In the third experiment, we test the accuracy of the proposed method under cyclic loading. The studied microstructure is shown in Figure 17(c) which has 40 pores and a pore volume fraction of 8.9%. In DNS, its domain is discretized by 283,596 linear tetrahedrons with 158,853 DOF. Two hardening laws are implemented for this experiment: isotropic hardening and linear kinematic hardening. We assume the microstructure starts from a relaxing initial state and is then subject to three sequential pure shear deformations ($0 \to F_1 \to F_2 \to F_3$) given in Equation (59):

$$F_1 = \begin{bmatrix} 1.0 & 0.005 & 0 \\ 0.005 & 1.0 & 0 \\ 0 & 0 & 1.0 \end{bmatrix}, \quad F_2 = \begin{bmatrix} 1.0 & -0.01 & 0 \\ -0.01 & 1.0 & 0 \\ 0 & 0 & 1.0 \end{bmatrix}, \quad F_3 = \begin{bmatrix} 1.0 & 0.015 & 0 \\ 0.015 & 1.0 & 0 \\ 0 & 0 & 1.0 \end{bmatrix} \quad (59)$$

The stress-strain behavior of the microstructure under loading-unloading-reloading is illustrated in Figure 21. As it can be observed, our predictions match with DNS with either linear or kinematic hardening laws.

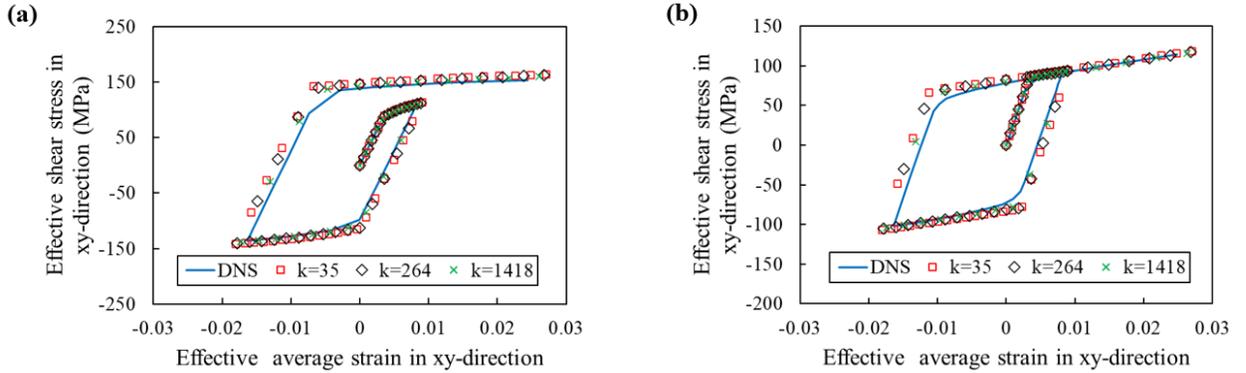

**Figure 21 Comparison of the homogenized stress components:** The comparison for the microstructure in Figure 16(c) is based on DNS and our ROM. Different hardening laws are used for validation: **(a)** Isotropic hardening. **(b)** Linear kinematic hardening.

Our ROM is designed for studying the influence of manufacturing-induced porosity on the hardening behaviors of cast alloys which typically have small pore volume fractions. However, to demonstrate our ROM's capability to simulate materials with high porosity, we use the deformation gradient in Equation (60) to deform a microstructure with a pore volume fraction of 15.9%.

$$F = \begin{bmatrix} 1.04 & 0 & 0 \\ 0 & 0.98 & 0 \\ 0 & 0 & 0.98 \end{bmatrix} \quad (60)$$

The morphology of the studied microstructure is shown in Figure 22(a) which has 25 pores. The aspect ratio and average nearest neighbor distance of pores are 1.4 and 24.3μm, respectively. We use this experiment to also demonstrate the advantages of our micro-ROM over FEA with a coarse mesh. We discretize this microstructure with two meshes: a sufficiently fine mesh with 68,675 tetrahedrons in Figure 22(b) and a coarse mesh containing 12,995 tetrahedrons in Figure 22(c). We consider the FE solutions of the fine mesh as DNS and provide Von-Mises stress distribution based on the fine and coarse meshes in Figure 23(a) and (b), respectively. In the case of ROM, we choose four clustering levels in the simulations, see Figure 23(c-f). By comparing the results between DNS and ROM in Figure 23, we can see that when cluster numbers are small, stress is mainly concentrated in locations where pores are closely packed and the contrast between low and high stress values is not as sharp as DNS. This is due to the averaging effects of clusters. When more clusters are used in ROM, both the stress distribution and the stress contrast show significant similarity to their counterparts in DNS. It is also noted that even though the FE coarse mesh has far more elements than our ROM (even when k=3200), its accuracy is much lower.

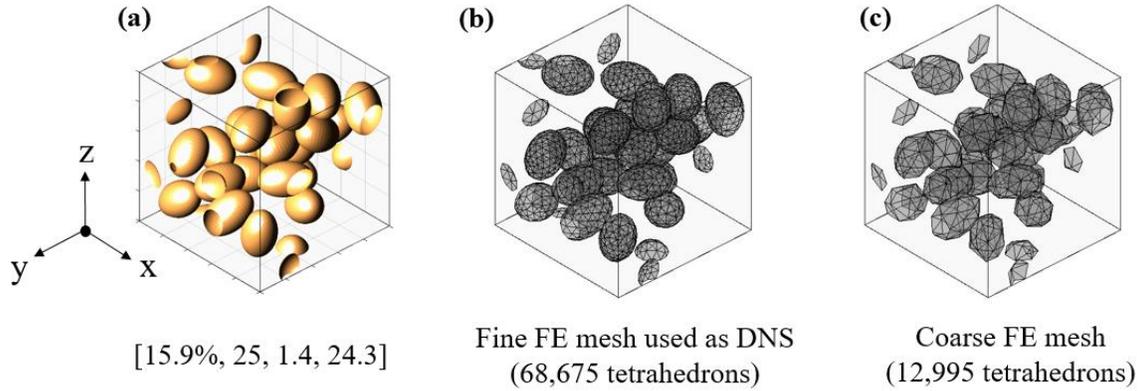

**Figure 22 Porous microstructure and discretization: (a)** Porosity morphology with pore descriptor values listed as $[V_f, N_p, A_r, \bar{r}_d]$. **(b)** A sufficiently fine FE mesh with 68,675 elements is used as DNS in this experiment. **(c)** The microstructure is also discretized by a coarse mesh with 12,995 elements where the shapes of the pores are not well represented. For clarification, only meshes on the pore surfaces are shown.

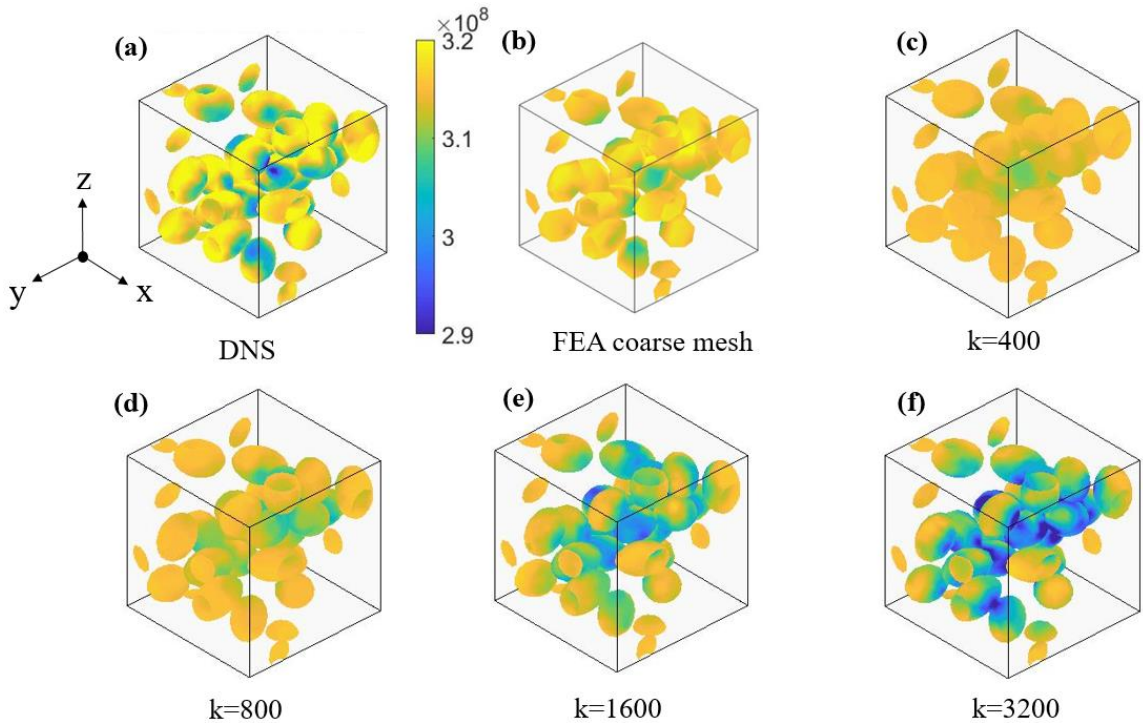

**Figure 23 Comparison of the Von-Mises stress distributions in microstructures: (a)** The Von-Mises stress distribution is obtained via DNS with sufficiently fine mesh. **(b)** The stress distribution is computed on a coarse FE mesh. **(c-f)** The stress distributions are approximated by ROM with different numbers of clusters.

To quantitatively assess the convergence of our approach in the case of a microstructure with a high pore volume fraction, we compare the homogenized stress-strain curves between DNS and ROM in Figure 24 and their associated toughness values in Table 1. Similar to the previous experiments, ROM's accuracy is improved when more clusters are utilized. However, different from scenarios with relatively low porosity volume fractions (e.g., in Figure 16), more clusters are required in this case to capture pore morphology and, in turn, match DNS. This observation is consistent with our modeling experience that prediction of the homogenized responses becomes more difficult as porosity volume fraction increases.

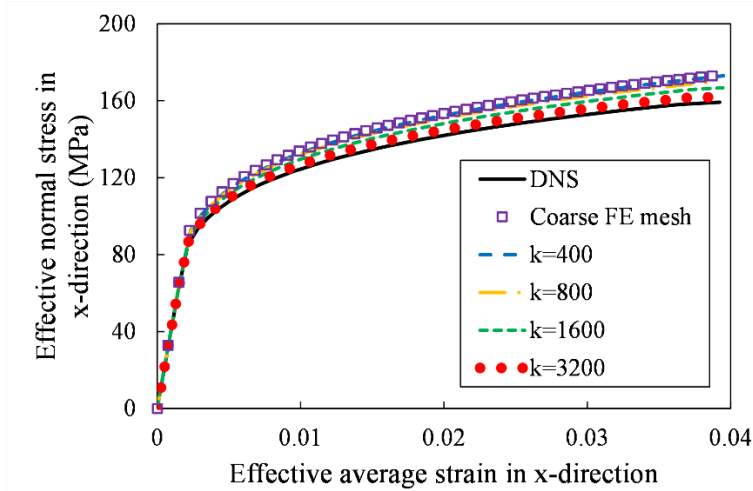

**Figure 24 The homogenized stress-strain curves:** The comparison is between DNS and ROM (with the different number of clusters) on the microstructure shown in Figure 23.

To demonstrate the acceleration effects of the proposed ROM, we compare its computational costs at different stages against DNS, see Table 2. As the number of clusters increases, both offline (clustering) as well as online (solution) costs increase and as a result the speedup factor decreases. In this example, the ROM's acceleration factor ranges from 4.2 to 72.7. In particular, with 3200 clusters our approach provides a 4.2 times speedup while having an error of 3.47% in predicting the toughness. We note that when calculating the speedup, the offline cost for cluster creations is not included as it is only done once and its results can be re-used for any deformation predictions. We note that our method does not require offline elastic tests and its offline stage only involves the generation of clusters and initialization of the related variables. Our cluster sizes are quite similar throughout the domains in this work and, as a result, more clusters (compared to SCA-like methods) are often required to capture local effects. We also point out that with the increase of clusters, the underlying algebraic system along with its numbers of eigenmodes also grow. With higher numbers of eigenmodes, clusters are able to approximate more sophisticated deformations with higher accuracy at higher costs.

**Table 1**: **Predicted toughness:** Comparison of toughness values simulated by DNS and ROM for microstructure analyses in Figure 23 and Figure 24.

| Method | Toughness (MJ/mm$^3$) | Error (%) |
| --- | --- | --- |
| DNS | 5.28 | - |
| ROM (k=400) | 5.71 | 7.53% |
| ROM (k=800) | 5.60 | 5.71% |
| ROM (k=1600) | 5.54 | 4.69% |
| ROM (k=3200) | 5.47 | 3.47% |

**Table 2**: **Computational time (unit: sec):** Comparison of different stages between DNS and ROM for microstructure analyses in Figure 23 and Figure 24.

| Method | Number of eigenmodes | Offline (clustering) | Online (solution) | Total | Speedup factor (DNS/online) |
|---|---|---|---|---|---|
| DNS | 41349 | - | 7288.8 | 7288.8 | - |
| ROM (k=400) | 2400 | 56.7 | 100.2 | 156.9 | 72.7 |
| ROM (k=800) | 4800 | 167.6 | 278.2 | 445.8 | 26.2 |
| ROM (k=1600) | 9600 | 317.7 | 633.3 | 951.0 | 11.5 |
| ROM (k=3200) | 19200 | 687.1 | 1738.2 | 2425.3 | 4.2 |

*5.3. Multiscale experiments*

In this section, we combine the two accelerating schemes as a multiscale ROM and compare its efficiency with the classic DNS (FE$^2$) approach. We use the 3D bracket in Figure 25 as the macrostructure. To reduce memory requirements, we assume pores only exist in the middle part of the bracket and only model one-quarter of the porous part due to the symmetric loading and boundary conditions, see Figure 25(a).

We study two multiscale models, as shown in Figure 25(b) and Figure 25(c). The purpose of the first model is to verify the accuracy of our multiscale ROM by comparing it with DNS. Due to its simple porosity morphology, we only need a relatively coarse FE mesh which results in acceptable computational costs for simulating elastoplastic responses. On the contrary, we integrate more complex porosity morphologies in the second model in Figure 25(c) to mimic actual pore distributions on a manufactured component to study the impact of spatially varying micromorphologies on the macrostructural behaviors. Since a much finer FE discretization is needed to capture the geometry details of local porosity morphology, computational expenses become prohibitively high for DNS. We thus only use our multiscale ROM in this case.

As shown in Figure 23(b), in the first multiscale model every macroscopic material point is associated with the same microstructure instance. This microstructure has one spherical pore in its center which accounts for a 6.5% pore volume fraction. For DNS, the two-scale model is discretized by 6.2 million finite elements in total, which includes 945 macroscale elements and 6,574 elements in each microstructure. For our ROM, the macrostructure is decomposed into 10 clusters and 157 clusters approximate each microstructure. Both DNS and ROM are implemented in MATLAB, which is run in parallel by 12 cores during runtime.

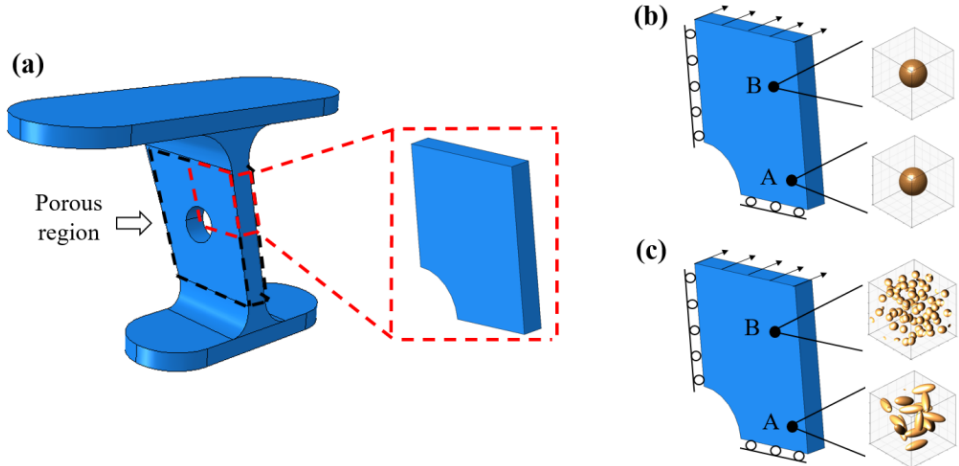

**Figure 25 Multiscale models: (a)** Pores are assumed only to exist in the middle section of the 3D bracket. Only a quarter of the middle part is modeled as the macrostructure in multiscale simulations. Two multiscale models are studied: **(b)** The first multiscale model with homogeneous porosity assigns each material point with an identical porous microstructure with a single spherical pore. **(c)** The second multiscale model with heterogeneous porosity assigns material points with microstructures of distinct porosity morphologies, as shown in Figure 9. In models **(b)** and **(c)**, all microstructures have the same pore volume fraction of 6.5%.

The two-scale Von-Mises stress distributions are compared between the DNS and ROM in Figure 26. Microstructural stress distributions are illustrated at two material points with different stress magnitudes along with macroscale stress distributions. We note that both the macroscopic and microscopic stress fields share significant similarities between the two methods. We also notice minor differences on local micro-stress fields where our ROM's local values appear smoother than DNS in a diffusive manner. We have discussed the same observation in Section 5.2.

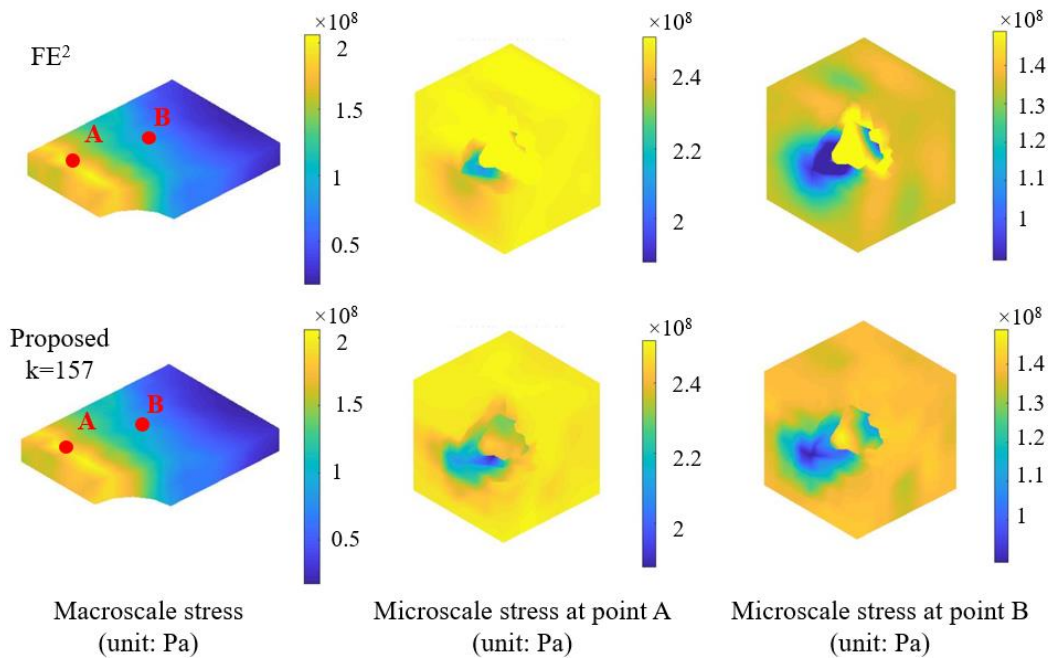

**Figure 26 Multiscale simulation with homogeneous porosity distribution:** Top row shows the DNS results of the Von-Mises stress distributions on the macro-structure and two microstructures at the material points A and B, respectively; the bottom row demonstrates the results from the proposed ROM.

To quantify solution differences between DNS and our ROM, we plot relative differences of the micro-stress fields at points A and B in Figure 27. The relative difference is computed by comparing pointwise Von-Mises stress values in the corresponding microstructures. From the two histograms, we find that the overall stress fields computed by the two methods agree well where most pointwise stresses from DNS overlap with their counterparts of ROM. To further quantify the difference, an L2-norm of the difference ($e$) of the pointwise Von-Mises stresses is computed:

$$e = \frac{1}{N_{ip}} \|\bar{\sigma}_{DNS} - \bar{\sigma}_{ROM}\|_2 \tag{61}$$

where $N_{ip}$ is the number of integration points in the microstructures, $\bar{\sigma}_{DNS}$ and $\bar{\sigma}_{ROM}$ are the Von-Mises stresses computed via DNS and ROM, respectively. The L2-norms of Figure 27(a) and Figure 27(b) are 0.044% and 0.084%, respectively, indicating very close stress distributions between the two methods. The relations between macroscopic reaction forces and tip displacements for the homogeneous porosity are demonstrated in Figure 28(a). We observe a generally good agreement between the two curves where their maximum difference is smaller than 3%. In terms of computational cost, the DNS took 528.1 hours, while our ROM is finished in 27.3 hours.

The second multiscale model aims to study the impacts of spatially varying porosity on structural behaviors. This model has the same macrostructure as the first one but it possesses spatially varying microstructures. Specifically, we randomly assign one of the microstructures in Figure 9 to each macro-point. Although each microstructure in Figure 9 has very different pore morphology and spatial distribution, its pore volume fraction (6.5%) is the same as the first model. In terms of domain discretization, the macro-domain is discretized by 945 elements and the microstructures in Figure 9(a)-(f) are meshed by 103,344, 123,552, 141,917, 153,815, 60,356 and 78,339 elements, respectively. In total, this multiscale model consists of 104.2 million elements. Since its DOF is approximately 17 times larger than the first model, the projected computational time of the DNS approach is about 8,875 hours, and hence we only use our ROM for this example. In our ROM, the macro-domain is decomposed by 10 clusters while each microstructure is discretized by 592 clusters. The simulation via our ROM is converged in 69.3 hours.

The Von-Mises stress distributions on both scales are illustrated in Figure 28(b)-(d). Even though the macrolevel stress distribution is similar to its counterparts in the first multiscale model, the stress distributions of the two microstructures are different from the ones in Figure 26 due to complex local porosity morphologies. We note the microscale stress values at the two points in this multiscale model appear higher than their counterparts in the first model with homogeneous porosity.

To quantify the effect of spatially varying pores on structural performance, the macroscopic reaction force and displacement of the heterogeneous porosity are compared between the two multiscale models, as shown in Figure 28(a). We notice two observations by comparing the two nonlinear curves. First, their elastic responses are close, since the two multiscale models share the same microstructural pore volume fraction (6.5%). The elastic behaviors seem to strongly depend on the pore volume fraction value, consistent with the observation reported in [66] where sensitivity analysis indicates particle volume fraction is the most influential geometry descriptor to determine the a composite's elastic responses. We also observe that the plastic reaction with heterogeneous porosity is noticeably (7.7%) higher than that with homogeneous porosity. One of the plausible reasons is that the heterogeneous porosity introduces higher stress concentrations at complex local morphologies, e.g., at points A and B. With higher microstructural stresses, the total macroscopic reaction force is more considerable.

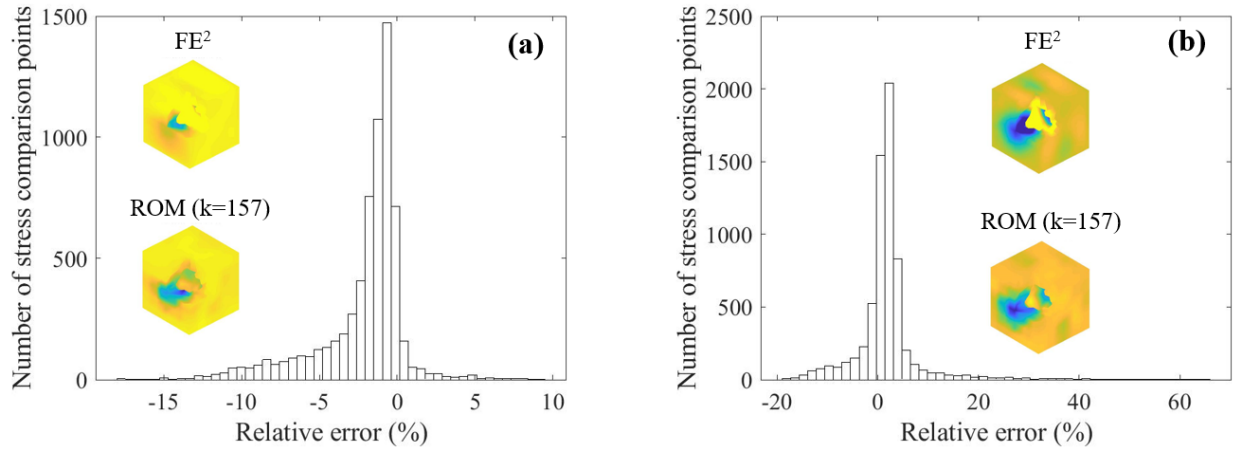

**Figure 27 Comparison of microscale stress fields between DNS (FE$^2$) and our ROM: (a)** Comparison of microscale Von-Mises stresses at point A, with an L2-norm difference of 0.044%. **(b)** Comparison of microscale stresses at point B, with an L2-norm difference of 0.084%.

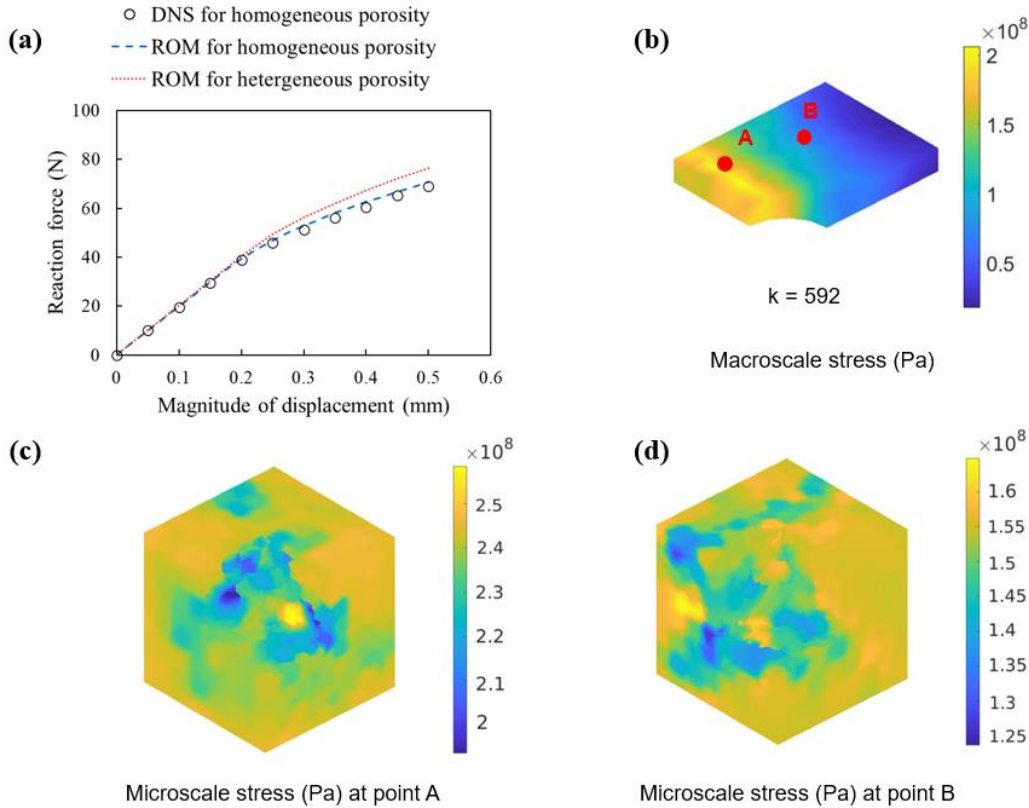

**Figure 28 Multiscale simulation results: (a)** Comparison of macroscale reaction force and tip displacement. The maximum difference for the homogeneous porosity simulations between DNS and ROM is smaller than 3%, while the reaction forces of the heterogeneous model is 7.7% higher than the homogeneous counterpart. **(b)** The Von-Mises stress distributions on the macro-structure. **(c)-(d)** The Von-Mises stress distributions in the microstructure associated with points A and B.

## 6. Conclusion

In this paper, we propose a new multiscale ROM, coined as deflated clustering analysis, to simulate the elastoplastic behaviors of heterogeneous alloys with complex microscopic pores. In particular, the proposed ROM consists of several significant components. First, we implement a spatial domain decomposition algorithm to significantly reduce the system's unknown variables from an FE mesh to a small group of clusters. The clustering process universally applies to both microscale and macroscale models by agglomerating nodes in proximity. Second, we accelerate the macroscale simulations by the incremental deflation method which is particularly useful for macrostructures with low plasticity percentages. While the macroscale acceleration scheme enhances computational efficiency by improving the CG solver's convergence and preventing unnecessary stiffness re-assembly during runtime, it ensures the solution accuracy of local deformations at macro-integration points. Third, we propose a microscopic projection method to model the nonlinear microstructural behaviors in a lower-dimensional space where reduced mesh and stiffness matrices are constructed to account for cluster interactions and strain evolutions. Fourth, we integrate a porosity-oriented microstructure characterization and reconstruction algorithm with the proposed ROM to mimic the local material heterogeneity caused by spatially varying porosity. In numerical experiments, we demonstrate that the proposed multiscale reduced model is highly accurate and computationally efficient.

Our ROM shares quite a few similarities with SCA but, contrary to SCA which groups elements based on their mechanical responses, our ROM agglomerates elements based on their geometrical proximity. It is has been shown [16–18,67] that the SCA-like methods perform reasonably well with a few clusters on composites or polymers. Although we have not done a one-on-one comparison, we think our ROM would need a few more clusters to represent the domain's topology and obtain solutions that are close to DNS. We would also like to point out that we find SCA to be commonly applied to composites with strong or weak inclusions where the property ratio between material phases (e.g., moduli) is reasonably small. Our ROM is used to simulate alloys with pores where the moduli difference between material and void is infinite. Compared to strong or weak inclusions, it is much harder to simulate microstructures with pores, especially for an FFT-based approach whose computational efficiency decreases as the phase contrast in a microstructure increases. The existence of pores is another reason that in this work we are reporting results based on a few more clusters compared to SCA when it is applied to composites. We note that porous materials are successfully modeled in a recent study [68] which suggests SCA-like methods may efficiently solve the porous models after modifications.

Our ROM has some major differences with a coarse-meshed FEM. Finite elements typically have similar geometry and shape (e.g., tetrahedral) and a coarse FE mesh may lack sufficient DOF to accurately represent high eigenmodes. Contrarily, clusters can be very different in shape. To model an irregular region in a geometry, FEM may need several elements while the ROM may only need one cluster. In addition, our clustering approach depends on the deflated CG which deflates the Krylov subspace with pre-defined cluster's rigid body modes and removes the smallest eigenvalues from the fine-meshed FEM. Since a CG's convergence mainly depends on the smallest eigenvalues, DCG can converge to the (fine-meshed) FEM solution in much fewer iterations. Therefore, compared to a coarse-meshed FEM, our clustering approach can converge to the accurate (fine-meshed) solutions more efficiently.

The proposed method can be improved in a few aspects. First, the reduced mesh is based on tetrahedrons generated by Delaunay triangulation. While tetrahedron mesh is advantageous in adapting to complex domain geometries, its computational accuracy could be problematic when its geometry is ill-shaped. A more robust tetrahedron-based meshing algorithm will increase the flexibility of our approach. Second, a node numbering algorithm needs to be introduced to reduce the bandwidth of the reduced stiffness matrix for improving matrix operation efficiency [53]. Third, only one-quarter of the porous part is simulated in the multiscale model in Section 5.3 to lower computer memory requirements. A feasible approach to reduce memory dependency is to utilize the assembly-free technique [46] where no global stiffness matrix is assembled. Fourth, in Section 5.3 we noted that pore morphological descriptors can play an essential role in determining plastic behaviors. To quantify the impacts of each descriptor on the plastic response of the material, a surrogate model can be fitted whose training data can be generated via our ROM. Finally, since our ROM is designed for metallic components with manufacturing induced pores, its performance on other material systems such as composites and ceramics needs further study.

## Acknowledgments

The authors thank the anonymous reviewers as well as the ACRC consortium members. Specifically, we appreciate Randy Beals from Magna International, and Chen Dai from VJ Technologies for providing us with the W-profile plate samples and X-ray computed tomography (CT) scanning and data generation, respectively. Ramin Bostanabad also acknowledges support from National Science Foundation (award number OAC-2103708).

[57] K. Meftah and L. Sedira, "A Four-Node Tetrahedral Finite Element Based on Space Fiber Rotation Concept," *Acta Materialia*, vol. 11, pp. 67–78, Dec. 2019, doi: 10.2478/auseme-2019-0006.
[58] R. D. Cook, D. S. Malkus, M. E. Plesha, and R. J. Witt, *Concepts and Applications of Finite Element Analysis, 4th Edition*, 4th Edition. New York, NY: Wiley, 2001.
[59] R. Bostanabad, "Reconstruction of 3D Microstructures from 2D Images via Transfer Learning," *Computer-Aided Design*, vol. 128, p. 102906, Nov. 2020, doi: 10.1016/j.cad.2020.102906.
[60] R. Bostanabad et al., "Uncertainty quantification in multiscale simulation of woven fiber composites," *Computer Methods in Applied Mechanics and Engineering*, vol. 338, pp. 506–532, Aug. 2018, doi: 10.1016/j.cma.2018.04.024.
[61] R. Bostanabad et al., "Computational microstructure characterization and reconstruction: Review of the state-of-the-art techniques," *Progress in Materials Science*, vol. 95, pp. 1–41, Jun. 2018, doi: 10.1016/j.pmatsci.2018.01.005.
[62] R. Bostanabad et al., "11 - Multiscale simulation of fiber composites with spatially varying uncertainties," in *Uncertainty Quantification in Multiscale Materials Modeling*, Y. Wang and D. L. McDowell, Eds. Woodhead Publishing, 2020, pp. 355–384. doi: 10.1016/B978-0-08-102941-1.00011-0.
[63] I. M. Sobol, "On quasi-Monte Carlo integrations," *Mathematics and Computers in Simulation*, vol. 47, no. 2, pp. 103–112, Aug. 1998, doi: 10.1016/S0378-4754(98)00096-2.
[64] H. Xu, Y. Li, C. Brinson, and W. Chen, "A Descriptor-Based Design Methodology for Developing Heterogeneous Microstructural Materials System," *Journal of Mechanical Design*, vol. 136, no. 5, Mar. 2014, doi: 10.1115/1.4026649.
[65] "ABAQUS/Standard User's Manual, Version 6.9. / Smith, Michael. Providence, RI : Dassault Systèmes Simulia Corp, 2009."
[66] M. A. Bessa et al., "A framework for data-driven analysis of materials under uncertainty: Countering the curse of dimensionality," *Computer Methods in Applied Mechanics and Engineering*, vol. 320, pp. 633–667, Jun. 2017, doi: 10.1016/j.cma.2017.03.037.
[67] Y. Yang, L. Zhang, and S. Tang, "A comparative study of cluster-based methods at finite strain," *Acta Mechanica Sinica*, p. 1, Aug. 2021, doi: 10.1007/s10409-021-01141-8.
[68] Y. Nie, Z. Li, and G. Cheng, "Efficient prediction of the effective nonlinear properties of porous material by FEM-Cluster based Analysis (FCA)," *Computer Methods in Applied Mechanics and Engineering*, vol. 383, p. 113921, Sep. 2021, doi: 10.1016/j.cma.2021.113921.
43